\documentclass[acmjacm,hyperref]{acmtrans2m}
\usepackage{amsmath}
\usepackage{graphicx}
\usepackage{verbatim}
\newtheorem{theorem}{Theorem}[section]

\newtheorem{proposition}[theorem]{Proposition}
\newtheorem{lemma}[theorem]{Lemma}
\newdef{definition}[theorem]{Definition}
\newdef{remark}[theorem]{Remark}

\newcommand{\eps}{\varepsilon}
\newcommand{\dsmall}{\delta_{\mathrm{small}}}
\newcommand{\dlarge}{\delta_{\mathrm{large}}}
\newcommand{\Wleft}{W_{\mathrm{left}}}
\newcommand{\Wright}{W_{\mathrm{right}}}
\newcommand{\eqdef}{:=}

\newif\iffull
\fulltrue
\newif\ifjournal

\title{Minimum-Weight Triangulation is NP-hard}

\author{WOLFGANG MULZER \\ Princeton University
        \and G{\"U}NTER ROTE \\ Freie Universit{\"a}t Berlin}

\begin{abstract}
A triangulation of a planar point set $S$ is a maximal plane 
straight-line graph with vertex set $S$.
In the \emph{minimum-weight triangulation} (MWT) problem, we are looking 
for a triangulation of a given point set that minimizes  the sum of the 
edge lengths. We prove that the decision version of this problem is 
NP-hard, using a reduction from PLANAR 1-IN-3-SAT. The correct 
working of the gadgets is established with computer assistance, using
dynamic programming on polygonal faces, as well as
the $\beta$-skeleton heuristic to certify that certain edges belong to
the minimum-weight triangulation.
\end{abstract}

\terms{Algorithms, Theory}
\keywords{Optimal triangulations, PLANAR 1-IN-3-SAT}

\category{F.2.2}{Nonnumerical Algorithms and Problems}
{Geometrical problems and computations}
\category{G.2.2}{Graph Theory}{Graph algorithms}

\begin{document}



\maketitle
\iffull
{\makeatletter
\gdef\@runningfoot{Technical Report B05-23 (revised), arXiv:cs/0601002v3,
February\ 2008}
\gdef\@firstfoot{Technical Report B05-23 (revised), arXiv:cs/0601002v3,
February\ 2008, \pageref{@lastpg} pages.}
}
\fi

\section{Introduction}

Given a set $S$ of points in the
Euclidean plane, a \emph{triangulation} $T$ of $S$ is a maximal plane 
straight-line graph with vertex set $S$. The \emph{weight} of $T$
is defined as the total Euclidean length of all edges in $T$. 
A triangulation that achieves the minimum
weight is called a \emph{minimum-weight triangulation} (MWT) of $S$. 

The problem of computing a triangulation for a given planar point set 
arises naturally in many applications such as stock cutting,
finite element analysis, terrain modeling, and numerical approximation.
The \emph{minimum-weight} triangulation
has attracted the attention of many researchers, mainly due to its natural
definition of optimality, and not so much because it would be important for
the mentioned applications.
We show that computing a {minimum-weight} triangulation is NP-hard. Note that
it is not known whether the MWT problem is in NP, because it is an open
problem whether sums of radicals (namely, Euclidean distances) can be compared 
in polynomial 
time \cite{Blomer91}. This problem is common to many geometric optimization
problems, like, perhaps most famously,  the Euclidean Traveling Salesperson 
Problem. To get a variant
of the problem that is in NP, one can take the weight of an edge $e$ as the
rounded value $\lceil \| e \|_2 \rceil$. Our proof also shows that this variant
is NP-complete.

Our proof uses a polynomial time reduction from POSITIVE PLANAR 1-IN-3-SAT,
a variant of the well-known PLANAR 3-SAT problem, which is a standard tool
for showing NP-hardness of geometric problems.

\subsection{Optimal Triangulations}

\begin{figure}
\begin{center}
\includegraphics{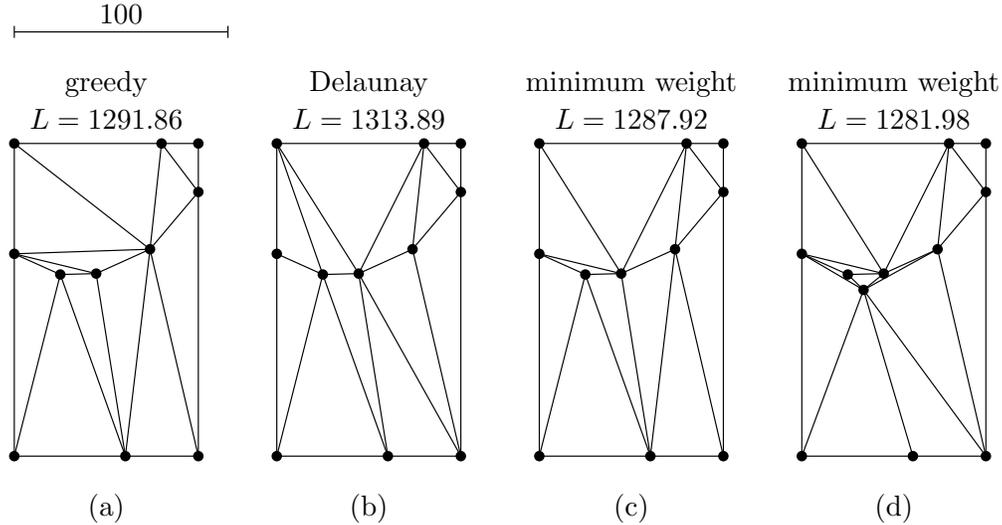}
\caption{A planar point set and different ways to triangulate it. The greedy 
triangulation (a) is constructed incrementally, always adding the shortest possible edge. In this example,
it is shorter than the Delaunay triangulation (b), which avoids skinny 
triangles. A minimum weight triangulation for the point set is shown in (c).
The length $L$ of the MWT can decrease when additional Steiner points are
allowed (d).}
\label{fig:triangulations}
\end{center}
\end{figure}

Usually, a planar point set has (exponentially) many different 
triangulations, and many applications call for triangulations with
 certain good properties (see Figure~\ref{fig:triangulations}). 
Optimal triangulations under various optimality criteria
were extensively surveyed by~\citeN{BernEp92}. 

The \emph{Delaunay triangulation} is perhaps the best known
triangulation. It arises as the dual of the
Voronoi diagram and can be computed optimally in $O(n\log n)$ time, using
classical techniques \cite{deBergKrOvSc2000}. Since it simultaneously 
optimizes many
objective functions (it maximizes the smallest angle, it minimizes the
maximum circumcircle,
as well as the maximum
smallest enclosing circle of all triangles), it is often the triangulation of choice.

Certain other criteria,
such as the minimum maximum angle,
the maximum minimum height, or the minimum maximum distance of a triangle
from its circumcenter,
can be optimized 
 in polynomial time using the
\emph{edge insertion} technique \cite{BernEdEpMiTa93,EdelsbrunnerTaWa92},
a sophisticated application of dynamic programming.

It makes a difference whether it is allowed to add new points, so-called 
\emph{Steiner points}, to the planar point set. As 
Figure~\ref{fig:triangulations} shows, this can sometimes help to improve
the objective function. \citeN{BernEp92} give many similar examples. Here we
only mention one result by \citeN{Eppstein94}, who showed that the
minimum weight \emph{Steiner} triangulation can be approximated up to a 
constant factor in $O(n \log n)$ time, using quadtrees.

\subsection{History of the MWT problem}

The minimum-weight triangulation problem dates back to the
1970s and has been called
``perhaps the most longstanding open problem in computational geometry''
\cite{BernEp92},
At the end of
the classical book of  \citeN{GareyJo79} 
on NP-completeness, there is a list of 12 major
problems whose complexity status was open at the time of writing. 
With the present paper, ten 
problems from this list have been resolved by proving NP-hardness or
by exhibiting a polynomial-time algorithm 
(see~\cite{Johnson05} for a recent status update on the list).
As of now, only two problems from the original list remain open, namely
Precedence-Constrained 3-Processor Scheduling and the notorious Graph
Isomorphism problem.

\subsubsection*{Early attempts}
It seems that the MWT problem was first considered by 
\citeN{DuppeGo70} who proposed a greedy algorithm 
which always adds the shortest possible edge to the triangulation.
Later, \citeN{ShamosHo75} suggested using the Delaunay triangulation
as a minimum-weight triangulation. \citeN{Lloyd77} provided 
examples which show that both proposed algorithms usually do not compute 
the MWT (Figure~\ref{fig:triangulations}). He also shows that it is 
NP-complete to decide whether the edge set of a given planar straight-line
graph (with crossing edges) contains a triangulation. After this, countless
researchers attacked the MWT problem from many different angles. In one line
of attack, researchers used classical optimization techniques such
as dynamic programming or branch and bound, but this soon became infeasible.
In other lines of research, people looked at relaxed variants of the problem:
Maybe there exist reasonable restrictions of the problem for
which efficient algorithms can be found, and if the problem cannot be solved
exactly, maybe good approximations can be computed efficiently. Finally, there
was an attempt to gain a better understanding of the geometric properties of
the MWT in order to find footholds for effective heuristics. We now describe
these approaches in more detail.

\subsubsection*{Dynamic Programming}
\citeN{Gilbert79} and \citeN{Klincsek80} independently showed how 
to compute a minimum-weight triangulation of a simple polygon in 
$O(n^3)$ time by dynamic programming. 
In fact, this problem has become one of the standard textbook examples 
(or exercises) for illustrating the dynamic programming paradigm.
There have also 
been attempts to attack the general problem with dynamic programming 
techniques. For example, \citeN{ChengGoTs95} 
used dynamic programming in order to compute a minimum-weight triangulation of 
a given point set $S$ in $O(n^{k+2})$ time if a subgraph of a MWT 
of $S$ with $k$ connected components is known. Using branch and cut, 
\citeN{KyodaImTaTa97} managed to compute MWTs of $100$ points, but for large
point sets mere dynamic programming becomes absolutely infeasible.

\subsubsection*{Restricted Instances}
For restricted classes of point sets, it is possible to compute the MWT in 
polynomial time. For example, \citeN{AnagnostouCo93} 
gave an algorithm to compute the MWT of the vertex set of $k$ nested convex 
polygons in $O(n^{3k+1})$ time. More recently, 
\citeN{HoffmannOk06} showed how to obtain the MWT of a point set with $k$ inner 
points in $O(6^k n^5 \log n)$ time. 

\subsubsection*{Approximations}
In another line of attack, researchers were looking for triangulations that
approximate the MWT. The Delaunay triangulation is not a good candidate,
since it may be longer by a factor of $\Omega(n)$ 
\cite{Kirkpatrick80,ManacherZo79}.  The greedy triangulation approximates the 
MWT by a factor of $\Theta(\sqrt{n})$ 
\cite{ManacherZo79,Levcopoulos87,LevcopoulosKr96}. \citeN{PlaistedHo87} showed
how to approximate the MWT up to a factor of 
$O(\log n)$ in $O(n^2 \log n)$ time. 
\citeN{LevcopoulosKr96} introduced quasi-greedy triangulations, which 
approximate the MWT within a constant factor. \citeN{RemySt06}
discovered an approximation scheme for MWT  that runs in
quasi-polynomial time:
for every fixed $\eps$,
it finds a $(1+ \eps)$-approximation in $n^{O(\log^8 n)}$ time.

\subsubsection*{Subgraphs and supergraphs} A different line of research tried to 
identify criteria to include or exclude certain edges of the MWT.
\citeN{Gilbert79} showed that the MWT always contains the shortest edge. 
\citeN{YangXuYo94} extended this result by proving that edges which
join mutual nearest neighbors are in the MWT. A larger subgraph of the MWT,
the \emph{$\beta$-skeleton}, was discovered by \citeN{Keil94}. We describe it 
in Section~\ref{sec:beta-skeleton}.
In practice, the $\beta$-skeleton has many connected components, and thus 
does not help much in computing a MWT. Often, the \textit{LMT-skeleton 
heuristic} described by \citeN{DickersonKeMo97} yields much
better results. It uses the simple fact that a MWT is locally optimal in the
sense that it cannot be improved by flipping the diagonals of a convex empty
quadrilateral in the point set. The LMT-skeleton made it feasible to compute
the MWT for larger, well-behaved point sets, and it has been the subject
of numerous further investigations
\cite{BeiroutiSn98,ChengKaSu96,AichholzerAuHa99,BellevilleKeMcSn96,BoseDeEv02}.

Approaching the problem from the other direction, 
\citeN{DasJo89} defined the diamond test, which yields a supergraph of the
MWT: An edge $e$ can only be contained in the MWT if at least one of the
two isosceles triangles with base $e$ and base angles $\pi/8$ is empty. This
constant was improved to $\pi/4.6$ \cite{DrysdaleMcSn01}
(see Figure~\ref{fig:beta-skeleton} below).
The diamond test gives an easy criterion to exclude impossible
edges from the MWT.
Usually, this eliminates all edges except a set of $O(n)$ remaining
candidate edges. (This statement is true for random point
sets, with high probability. With bucketing techniques, such a set of
$O(n)$ edges can be found in linear expected time~\cite{DrysdaleRoAi95}.)

\subsection{Our Methods and Results}

\subsubsection{Historical Perspective}\label{sec:perspective}

The crucial necessary condition for a geometric optimization problem
to be NP-hard is that the solution depends non-locally on the data.

\begin{figure}[htb]
  \centering
  \includegraphics{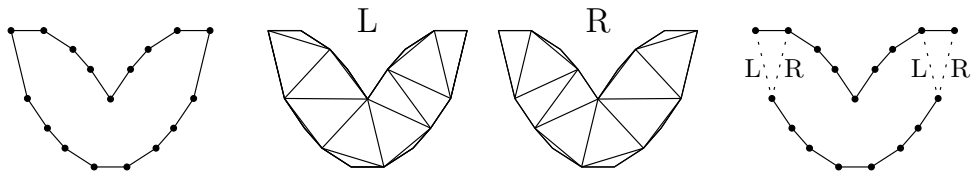}
  \caption{A wire-piece and its two optimal triangulations.  The two
    optimal triangulations are labeled as L and R, depending on
    whether the left (L) or right (R) arms of the isosceles terminal
    triangles, as shown dotted in the right
    part, belong to the triangulation. This convention is used
    throughout the paper.}
  \label{figure-wire}
\end{figure}

With the LMT-skeleton heuristic, it became feasible to compute
minimum-weight triangulations fast enough that one could carry out
experiments and play with various point sets.  An instance of a
non-local effect was hence discovered by Jack Snoeyink
\cite{BeiroutiSn98}: the so-called \emph{wire} is a symmetric polygon
 that does not have a symmetric minimum-weight triangulation, and
hence it has (at least) two different minimum-weight triangulations,
see Figure~\ref{figure-wire}.

\begin{figure}[htb]
  \centering
  \includegraphics{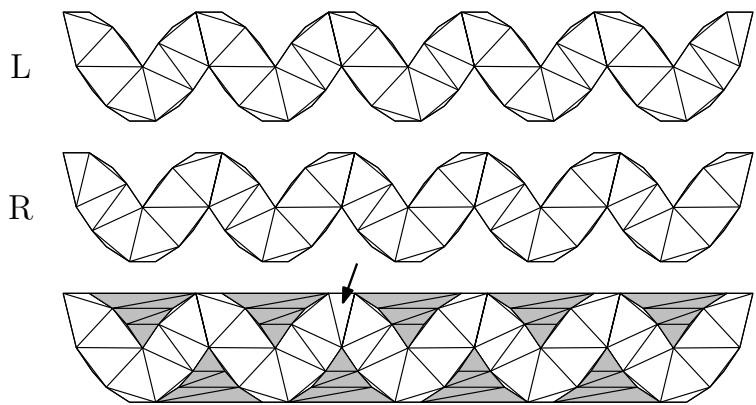}
  \caption{A longer wire with several optimal triangulations.  Besides
    the two triangulations of type L and R, there are mixed forms
    where the transition between L and R can occur in intermediate
    positions, as marked by an arrow in the lower example. If the wire
    is regarded as a point set, the shaded pockets outside the wire
    also have to be triangulated. For each pocket, there are four
    optimal triangulations. However, the pockets are independent of
    each other, and do not affect the properties of the wire.}
  \label{figure-long-wire}
\end{figure}

Wires can be extended to any length,
see Figure~\ref{figure-long-wire}.
\emph{A small change in the point set
at one end of the wire will cause the optimum triangulation to topple
globally throughout the whole wire.}
Wires can thus be used to
``transmit information'' from one area of the plane to a remote area
(hence the name ``wire'').  We use various forms of wires as
building blocks for constructing gadgets in our NP-hardness reduction.

Such a non-local effect is usually a strong indication that the
problem is NP-hard.
Still,
%
it took almost a decade until another crucial building block was
designed: a gadget that allows several wires to meet and that carries
out ``logic'' with the information that is transmitted by them (the
so-called $C$-connection, see Figure~\ref{C-connection}).  The design
of this gadget heavily depended on computer assistance.
We used an implementation of the LMT-skeleton heuristic
using Otfried Cheong's program \textsc{Ipe} \cite{Schwarzkopf95} as a 
graphical user-interface.

\subsubsection{Computer Assistance}

It seems inevitable that a NP-hardness proof for the
MWT problem requires \emph{some} amount of computer calculation, since
computing the weight of a triangulation involves
Euclidean distances, and thus square roots. Therefore, when comparing two
triangulations that differ in more than a pair of edges, it is
hard to compare their weight by just looking at them. 

However, since the conference version of this paper \cite{MulzerRo06},
we have made an effort to reduce the part of the proof that depends on
computer assistance for verification.
We have also made other simplifications of the proof and made it more accessible.
(a) We have tried to simplify and specialize the problem as much as
possible already at the logical (discrete) level.  In particular, the
PLANAR 1-IN-3-SAT problem is more specialized than in
 \cite{MulzerRo06} 
and does not need negations.
(b) We designed the gadgets in such a way that the $\beta$-skeleton
edges already form large connected components. This part of the construction
is trivial to check by computer, and is open even to visual inspection.
(c) In the proof, the gadgets split into modules (``pieces'') that can
be analyzed separately.
(d) The analysis of each module boils down to checking a small number
of possibilities, each amounting to computing the minimum-weight
triangulation of a simple polygon.
This can be programmed from scratch in less than an hour's work by an experienced programmer.
(e) To make the computer-assisted part even safer, we have rerun the final
calculations with interval arithmetic, using the open-source high-level
programming language \textsc{Python}, which automatically supports
integer arithmetic of unbounded length.

The final computer runs to verify our gadgets took about 10 hours on a 
relatively slow computer.
\iffull
The code is given in Appendix~\ref{app:programs} and amounts to approximately 1000 lines.
\fi

\section{Overview of the Reduction}
\label{sec:overview}

\begin{figure}
\begin{center}
\includegraphics[scale=1]{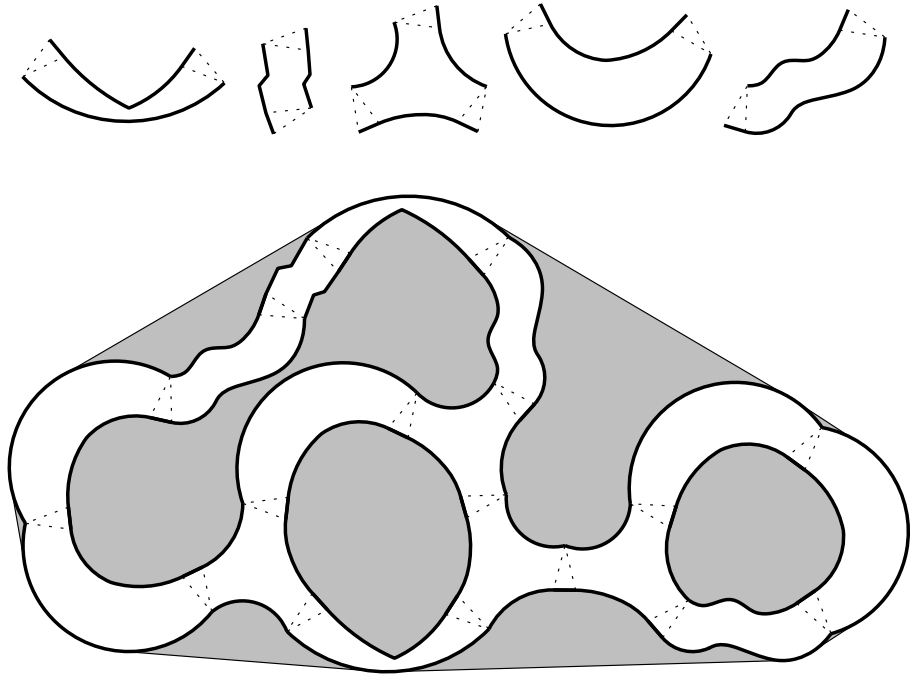}
\end{center}
\caption{A schematic view of the construction: From the five types of
  building blocks shown at the top, one can build a network of tubes
  like in the lower part. The area within this network is the
  interesting part where the simulation of the logic takes place. The
  \emph{holes} inside the structure and the \emph{pockets} between the
  structure and the convex hull (drawn shaded) form simple polygons
  that can be optimally triangulated in polynomial time.}
\label{fig:schematic}
\end{figure}

We begin with a high-level overview of the geometric construction.  We build our
point set from a small number of tube-like \emph{pieces} that fit
together at the openings, where they share \emph{terminal triangles}, see
Figure~\ref{fig:schematic}.  The boundary edges of the pieces must
belong to every optimal triangulation, regardless of how we put the
pieces together (Proposition~\ref{boundary} below). If we draw these edges we 
get a system of tubes, separated by polygonal holes.
The holes can be triangulated in polynomial time 
by dynamic programming; The tubes will form the essential part of the gadgets 
that model the logical structure of a POSITIVE PLANAR 1-IN-3-SAT formula,
whose satisfiability is NP-complete to decide, see Section~\ref{subsec:P1I3SAT}.

For the analysis, we will show that at each terminal triangle, we have
a choice of two edges to insert (Lemma~\ref{lemma-two-edges}).  When
one of the edges is inserted for every terminal triangle, the edges
enclose a simple polygon, whose optimum solution can be computed by
dynamic programming. Every piece can therefore be analyzed in
isolation, by considering a small number of possibilities.

\section{POSITIVE PLANAR 1-IN-3-SAT}
\label{subsec:P1I3SAT}

In this section, we describe the POSITIVE PLANAR 1-IN-3-SAT problem, which we 
will use for our reduction.

\begin{definition}
Let $\Phi$ be a Boolean formula in 3-CNF. The \emph{associated graph} of 
$\Phi$, $G(\Phi)$, has one vertex $v_x$ for each variable $x$ in 
$\Phi$ and one vertex $v_C$ for each clause $C$ in $\Phi$. There is an edge 
between a variable-vertex $v_x$ and a clause-vertex $v_C$ 
if and only if $x$ or $\neg x$ appears in $C$. 

The Boolean formula $\Phi$ is called \emph{planar} if its associated
graph $G(\Phi)$  is planar.
\end{definition}

\begin{figure}
\begin{center}
\includegraphics[scale=1]{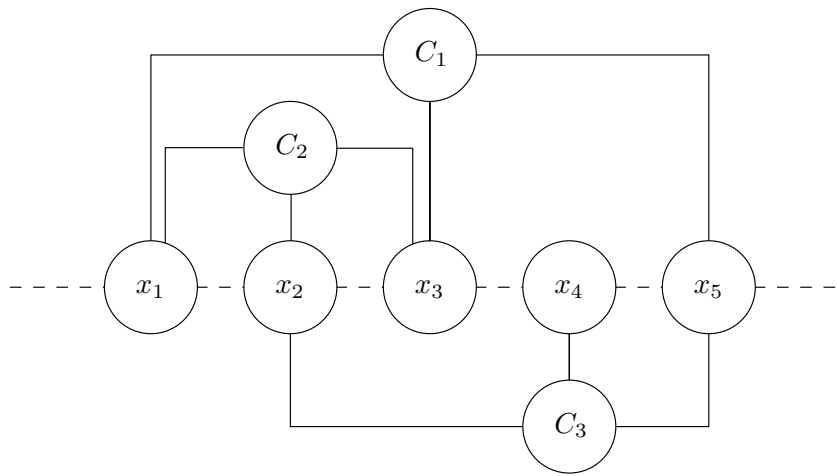}
\end{center}
\caption{A rectilinear embedding of graph that is associated with the
Boolean formula 
$\left(x_1 \vee \neg x_3 \vee x_5 \right) 
\wedge \left(\neg x_1 \vee x_2 \vee x_3 \right) \wedge 
\left(x_2 \vee x_4 \vee \neg x_5 \right)$.}
\label{fig:embed}
\end{figure}

\citeN{Lichtenstein82} showed that 3-SAT remains NP-complete
if the input is restricted to a planar formula
(the {PLANAR 3-SAT} problem).
As \citeN{KnuthRa92} observed, 
Lichtenstein's proof implies that it suffices to consider formulae 
whose associated graph can be embedded such that the variables are arranged on 
a straight line, with three-legged clauses above and below them. The edges 
between the variables and the clauses are embedded in a rectilinear fashion 
(see Figure~\ref{fig:embed}).

In our reduction we will use a variant of PLANAR 3-SAT
in which there are only positive variables and we ask for an assignment to the
variables such that in each clause exactly one variable is set to true.

\begin{definition}
  In the \emph{POSITIVE PLANAR 1-IN-3-SAT} problem, we are given a collection $\Phi$ of
  clauses containing exactly three variables together with a 
  planar embedding of the associated graph $G(\Phi)$ as described above.
  
  The problem is to decide whether there exists an assignment of truth values
  to the variables of $\Phi$ such that exactly one variable in each clause is
  true.
\end{definition}

The POSITIVE 1-IN-3-SAT problem without the planarity restriction was
shown to be NP-complete by \citeN{Schaefer78}.  This problem can also
be interpreted as a SET PARTITIONING problem, where every element of
the ground set has precisely three sets by which it can be covered.
For completeness, we include a proof that the planar version of the
problem is NP-complete.

\begin{proposition}
 POSITIVE PLANAR 1-IN-3-SAT is NP-complete.
\end{proposition}

\begin{proof}
It is easy to verify in polynomial time that a given embedding of a formula
is rectilinear (the validity of an instance) and that a given assignment has 
the 1-IN-3 property (the validity of a certificate). Hence,
the problem is in NP. To show completeness, we
describe a reduction from PLANAR 3-SAT. Let $I$ be an instance of 
PLANAR 3-SAT, i.e. a 3-CNF formula $\Phi$ and a planar, rectilinear 
embedding of the associated graph. We describe how to transform $I$ into
an instance of POSITIVE PLANAR 1-IN-3-SAT while maintaining the rectilinear
embedding. We consider the clauses of $\Phi$ one by one. 

If clause $C$ contains only one literal, it can easily be eliminated.
If clause $C$ contains two literals, say $x$ and $y$, we can
replace it by two three-variable clauses $(x \vee y \vee z)$ and
$(x \vee y \vee \neg z)$, where $z$ is a new variable.

Before we consider clauses with three variables, we first discuss two
useful gadgets which enforce equality and inequality between variables
in terms of 1-IN-3 clauses.

\begin{figure}[ht]
\begin{center}

\includegraphics[scale=1]{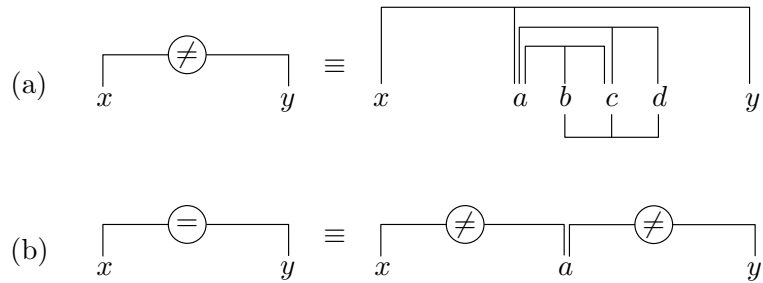}

\end{center}
\caption{The gadgets to enforce inequality (a) and equality (b) between two 
variables.}
\label{fig:negequ}
\end{figure}

\subsubsection*{The inequality gadget} The inequality gadget serves to enforce the 
constraint $x \Leftrightarrow \neg y$ for two variables $x$ and $y$. We abbreviate it
as $(x \neq y)$. It is implemented as 
\begin{align}
\label{ali:neg}
x \neq y \equiv (x, a, y) \wedge (a, b, c) \wedge (a, c, d)
\wedge (b, c, d), 
\end{align}
where $a$, $b$, $c$, $d$ are new variables used only inside the gadget (see
Figure~\ref{fig:negequ}a).
Here, as always, we denote a 1-IN-3 clause by a triple $(v_1, v_2, v_3)$.
The clause $(v_1, v_2, v_3)$ is \emph{satisfied} if exactly one of the variables
$v_1$, $v_2$, $v_3$ is true. 
\begin{lemma}
Given $x$ and $y$, the expression \thetag{\ref{ali:neg}} is satisfiable iff 
exactly one of $x$ and $y$ is true.
\end{lemma}
\begin{proof}
The last three clauses enforce that $a = 0$, because otherwise we get 
$b=c=d=0$, and the last clause is not satisfied. The first clause now ensures
that exactly one of $x$ and $y$ is true.
\end{proof}

\subsubsection*{The equality gadget} Using two copies of the inequality gadget and an
extra variable $a$, we can build a gadget to enforce the constraint $x \Leftrightarrow y$
(see Figure~\ref{fig:negequ}b). We abbreviate it as $(x = y)$.

\begin{figure}[ht]
\begin{center}

\includegraphics[scale=1]{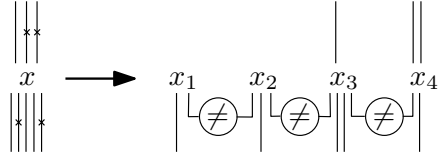}

\end{center}
\caption{Eliminating negations: The variable $x$ in this example
  has eight connections, some of which are negated, as indicated by
  crosses. We replace $x$ by a chain of variables $x_1,x_2, \ldots$ with
  alternating truth values and place the connections to the clauses
  containing $x$ correspondingly.}
\label{fig:repl3}
\end{figure}
\subsubsection*{Elimination of negated variables}
 To eliminate the negated variables in the original 3-SAT clauses, we
replace a variable $x$ by a chain of variables $x_1, x_2, \ldots$ and
use the negation gadget to enforce the appropriate relations between them
(see Figure~\ref{fig:repl3}).

\subsubsection*{Transformation of disjunctions}
It remains
 to handle a disjunctive clause $C=(x \vee y \vee z)$ with three 
literals $x$, $y$, and $z$. $C$ can be replaced by
\begin{align}
\label{1-in-3}
\left(x, u, a \right) \wedge \left( y, u, b\right) &\wedge
\left( a, b,  q \right) \wedge 
\left(u = c \right) \wedge
\left(d \neq z \right) 
\wedge
\left( c, d, r\right). 
\end{align}
See Figure~\ref{fig:repl2}. Note that there is space to accommodate 
the middle legs of the equality and inequality gadgets.

\begin{lemma}
Given $x$, $y$, and $z$,
the expression \thetag{\ref{1-in-3}} is satisfiable iff
$x \vee y \vee z$ holds.
\end{lemma}
\begin{proof}
The clause $( c, d, r)$ is equivalent to
$\neg c \lor \neg d$, since the variable $r$ appears nowhere else. Hence
$c\rightarrow \neg d \iff u\rightarrow z$ by the third and fourth 
constraint.
Thus, satisfiability of the last three constraints is equivalent to
$u\rightarrow z$.

Now, if $u=1$, the first three clauses can be satisfied only by
$x=y=0$, and we must have $z=1$.
The third clause is satisfied by setting $a=b=0$ and $q=1$.
For $u=0$, the first two clauses reduce to $x\ne a$ and $y\ne b$. 
Since $q$ appears nowhere else,
the clause $( a, b,  q)$ is equivalent to
$\neg a \lor \neg b \iff x\lor y$.
The value of $z$ can be arbitrary in this case.
\end{proof}

To make sure that $x$ and $z$ remain reachable from other clauses, we also 
add two additional variables $x'$ and $z'$ and equality 
constraints.
Clauses that were above the variables and nested between $x$ and $y$ can now 
connect to $x'$ instead of $x$, and those
nested between $y$ and $z$ can connect to $z'$ instead of $z$.

\begin{figure}[ht]
\begin{center}
\includegraphics[scale=1]{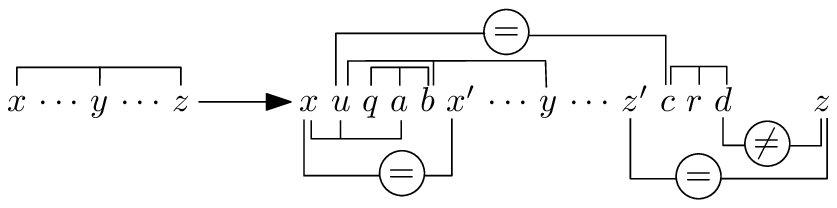}

\end{center}
\caption{The modification for a clause $x \vee y \vee z$ with three
  literals.}
\label{fig:repl2}
\end{figure}

This transformation can be carried out in polynomial time, and
by construction, the transformed positive formula has a rectilinear
embedding, and it has a 1-IN-3 assignment if and only if the original formula is 
satisfiable.
\end{proof}

\section{Geometric Validation Tools}

We summarize the geometric and computational tools we need for
establishing properties of the gadgets and their
minimum-weight triangulations. 

\subsection{The $\beta$-Skeleton}
\label{sec:beta-skeleton}

\begin{figure}[htb]
  \centering
  \includegraphics{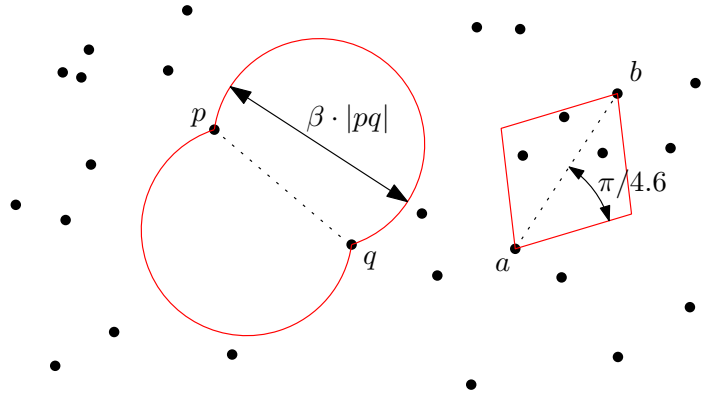}
  \caption{An edge $pq$ of the $\beta$-skeleton. The edge $ab$ does
    not
belong the MWT since the diamond test is violated.}
  \label{fig:beta-skeleton}
\end{figure}

The \emph{$\beta$-skeleton} of a point set $S$ is defined as the set of
all edges $pq$ between two points of $S$
such that the two circles of diameter
$\beta\cdot|pq|$ passing through $p$ and $q$ 
are empty, see Figure~\ref{fig:beta-skeleton}. It provides
a sufficient condition for including edges in the minimum-weight triangulation:

\begin{theorem}
\label{prop:beta}
The $\beta$-skeleton is a subgraph of the MWT for
\begin{displaymath}
\beta = \sqrt{1+\sqrt{4/27}} 
\leq 1.17682.
\end{displaymath}
\qed
\end{theorem}

Theorem~\ref{prop:beta} was first proven by
\citeN{Keil94} for $\beta \leq \sqrt{2}$ and was later improved by
 \citeN{ChengXu96}. The value for $\beta$ is nearly optimal,
since there is a lower bound of 
$\sqrt{5/4+\sqrt{1/108}} 
 \approx 1.16027$
\cite{WangYa01}.
In Proposition~\ref{boundary}, we will use the $\beta$-skeleton to identify the 
\emph{boundary}
edges of our gadgets, which must belong to every optimal triangulation.
Since the $\beta$-skeleton is defined by a local condition, it is very easy
to verify that a certain edge is a boundary edge. Our gadgets have many
points, hence we used a computer to check the $\beta$-skeleton property of
our boundary edges.
\iffull
The source-code of the straightforward program can be found in 
Appendix~\ref{app:beta-skeleton}.
\fi

\subsection{Triangulating a Polygon by Dynamic Programming}
\label{sec:dynamic-programming}

As we explained above, the analysis of our reduction essentially boils down
to computing the minimum-weight triangulation of a small number of simple
polygons. \citeN{Gilbert79} and \citeN{Klincsek80} independently showed how
to do this:
\begin{proposition}\label{prop:dyn}
The minimum-weight triangulation of a simple polygon can be computed in
$O(n^3)$ time.\qed
\end{proposition}

Proposition~\ref{prop:dyn} is proved by a straight-forward application of
dynamic programming, and we implemented a variant of the algorithm in 
\textsc{Python}.
\iffull
(The source code of our implementation is given in Appendix~\ref{app:dynprog}.)
\fi
The dynamic programming algorithm assumes an idealized model
in which arithmetic with edge lengths can be done exactly.
This means that
sums of radicals (i.~e.\ Euclidean distances) 
need to be compared in constant 
time. 
However, it is not known how to do this even in polynomial time~\cite{Blomer91}.  Therefore, our program uses interval arithmetic
to compare sums of distances, and for all our gadgets it computes an 
unambiguous answer.

Our program identifies the edges in a 
minimum-weight triangulation $T$ inside a given simple polygon $P$. We call the 
total weight of these edges the \emph{internal cost} of $P$ and denote it by 
$c(T)$. The internal costs computed by our program will be discussed 
extensively in Section~\ref{piece-analysis}.
It is easy to modify the program so that it computes the optimum
under the restriction that certain edges are forbidden. This is needed
for the proof of Lemma~\ref{lemma-two-edges}.

\section{Terminal Triangles}

\begin{figure}
  \centering
  \includegraphics[scale=0.8]
  {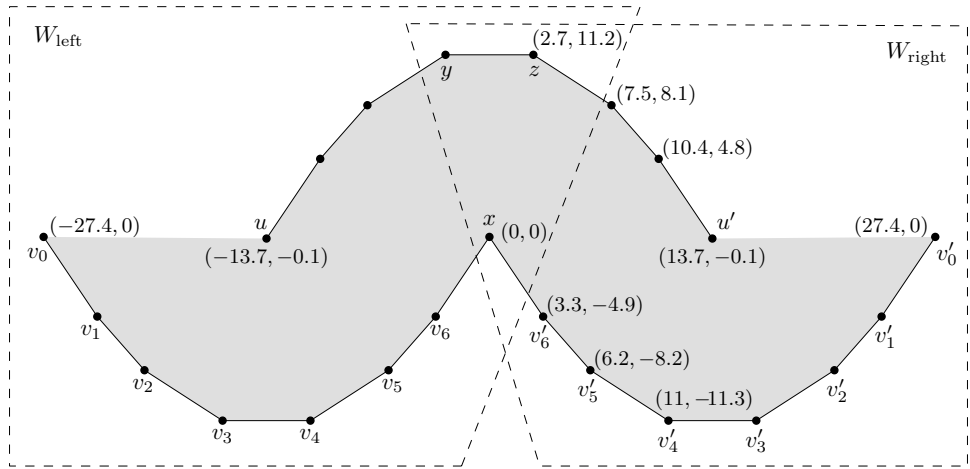}
    \caption{The point set $W$ in Lemma~\ref{lemma-two-edges}}
  \label{set-W}
\end{figure}

The point set is constructed from small \emph{elementary pieces} that fit
together at \emph{terminal triangles}. These terminal triangles are
isosceles triangles $xyz$ with base $yz$, that lie symmetrically with
respect to a coordinate axis, and they come in two 
sizes (small
and large). The coordinates of 
vertical terminal triangles are given
in Table~\ref{tab:terminal}. See Figure~\ref{set-W}
or~\ref{definition-wire}\iffull--\ref{definition-wire-extended}\fi\
  for an illustration.
\begin{table}[htb]
  \centering
\noindent\vbox{
\offinterlineskip
\halign{\vrule\strut\ #\ \hfil\vrule
 &\ (\hfil $#$.&$#$\hfil, &\hfil $#$.&$#$\hfil) \vrule
 &\ (\hfil $#$.&$#$\hfil, &\hfil $#$.&$#$\hfil) \vrule
 &\ (\hfil $#$.&$#$\hfil, &\hfil $#$.&$#$\hfil) \vrule
 &\ \hfil $#$ \vrule
\cr
\noalign{\hrule}
& \multispan4\hfil $x$\hfil \vrule
& \multispan4\hfil $y$\hfil \vrule
& \multispan4\hfil $z$\hfil \vrule
&\delta\hfil
\cr 
\hline
small & 0&00 & 0&00 &  -2&7 & 11&2 &   2&7 & 11&2 & \dsmall=\hphantom{9}5.655172\cr
large & 0&00 & 0&00 & -11&61 & 48&16 &  11&61 & 48&16 & \dlarge=
24.06\hphantom{9900}\cr
\hline
}}
  \caption{Terminal triangles. The ``difference terms'' $\delta$ will
    described below in
Section~\ref{sec:red-weight}.}
  \label{tab:terminal}
\end{table}
These triangles can be rotated by multiples of $90^\circ$, and
translated by multiples of 0.01 in each coordinate.  The large
triangle is a scaled copy of the small triangle, by a factor of~$4.3$.

The basic pieces that we construct are point sets with two or three
terminal triangles.  The $\beta$-skeleton edges will form a simple
polygon through these points, with a missing edge at each terminal
triangle.  We will show that in each terminal triangle, the edge $xy$
or the edge $xz$ must be present in every minimum-weight triangulation
(Lemma~\ref{lemma-two-edges}).  Thus, in order to determine how a piece with 
$k$ terminals can be triangulated in a minimum-weight triangulation, it suffices 
to consider $2^k$ possibilities for the positions of each
terminal edge. For each possibility, we find the optimum solution by
dynamic programming.


Let $W$ be the point set given in Figure~\ref{set-W}, which forms
part of the wire mentioned in the introduction 
(see Section~\ref{sec:perspective}).
  It is symmetric about the $y$-axis. The points $v_0$, $x$,
and $v'_0$ lie on the $x$-axis, and the points $u$ and $u'$ lie $0.1$
units below this line. The left and right halves of the lower boundary
are equal, and they are equal to the upper boundary (turned by
$180^\circ$).
We denote by $\Wleft$
and  $\Wright$ the left half and the right half of the set $W$,
up to and including the points $x$, $y$, and~$z$.

\begin{lemma}\label{lemma-two-edges}
  Let $P$ be a point set that contains a \textup(translated, scaled, rotated\textup)
  copy of the point set $W$, but no other points in the shaded polygon
  of Figure~\ref{set-W}.  Any minimum-weight triangulation $T$ of $P$
  that contains all edges shown in Figure~\ref{set-W} must contain at
  least one of the edges $xy$ or $xz$.
\end{lemma}
\begin{proof}
  The proof distinguishes 21 cases and deals with each case by a
  small calculation.

  Since in any triangulation, the point $u$ must be incident to some
  edge that emanates into the lower half-plane, $T$ must contain at
  least one of the edges $uv_1$, $uv_2$, \dots, $uv_6$.  Similarly,
  $T$ must contain at least one of the edges $u'v'_1$, $u'v'_2$,
  \dots, $u'v'_6$.
  Now, for each $i,j=1,\ldots,6$, the edges $uv_i$ and $u'v'_j$
  enclose a simple polygon. For each polygon, we calculated, by
  dynamic programming, the optimum triangulation as well as the
  optimum triangulation that uses none of the edges $xy$ and $xz$.
  By symmetry, we have to consider only $1\le i\le j\le 6$.
\iffull
  The results are shown in Table~\ref{W-lemma}.
\fi
\end{proof}

\begin{table}
\centering
\iffull
 \includegraphics[width=\textwidth] {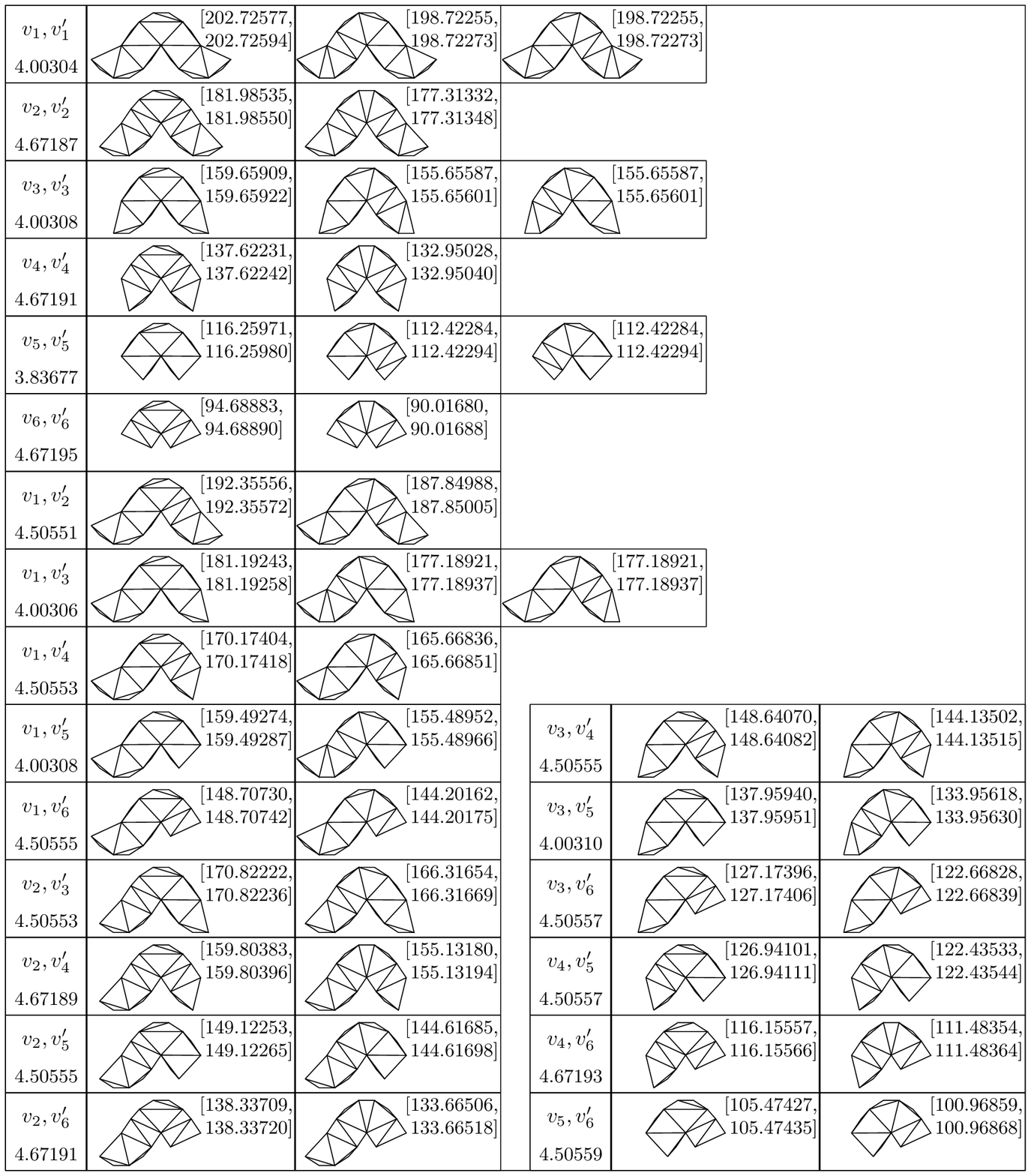}
\fi
  \caption
  {The optimum triangulations with and without the edges $xy$ and
    $xz$,
\iffull
 for each polygon bounded by 
\fi
edges $uv_i$ and $u'v_j'$, as
    indicated in the left column.  The second column shows an optimal
    triangulation without the edges $xy$ and $xz$, together with the
    interval for its internal cost, calculated to 5 digits after the
    decimal point.  (There are always at least two
    optimal triangulations, since the upper diagonal in the middle can
    be flipped. The program has picked one solution whose lower interval
    bound is smallest.)  The remaining columns list all
    candidates for optimal solutions.  One can check that all these
    solutions contain the edges $xy$ and $xz$.  (In fact, one can
    see that, if there are two optimal solutions in a row, they must
    have exactly the same weight since they contain parts which can be
    replaced by symmetric parts.)
    The lower bound on the difference between the optimum in the third
    column and the optimum in the second column is given in the first
    column.  The smallest difference ($3.83677$) occurs for the case
$v_5,v_5'$.
\iffull
The numbers in this table were computed with the \textsc{Python} program
in Appendix~\ref{app:two-edges}.
\fi
}
  \label{W-lemma}
\end{table}

\begin{figure}
  \centering
  \includegraphics[scale=0.8] {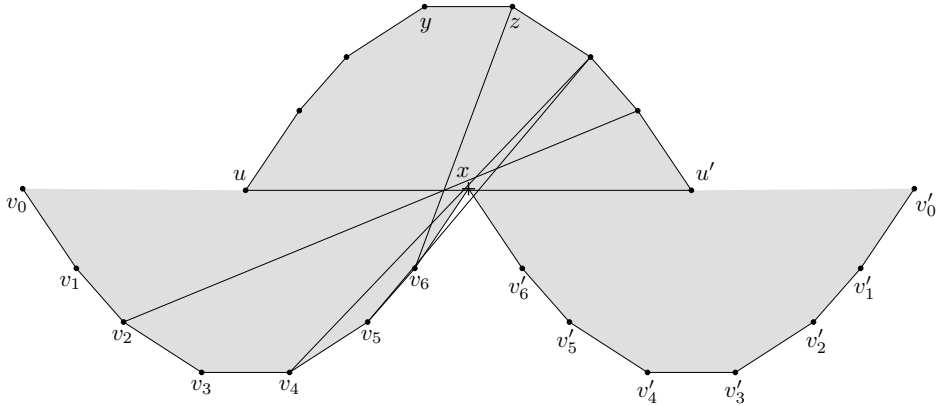}
  \caption{The point set $W$ in Lemma~\ref{lemma-two-edges} is in
    sufficiently general position.}
  \label{set-W-lines}
\end{figure}

We remark that the lemma is somewhat robust against perturbations and remains true even if all points of $W$ are moved by
a small distance.
Figure~\ref{set-W-lines} shows that small perturbations do not make
triangulations feasible or infeasible.
%
%
  The triangulations in Table~\ref{W-lemma} contain at most 18 internal edges.
  If each point is moved by at most $\sigma := 0.04$,
  the length of each edge changes by at most $2\sigma$, and the weight
  of any triangulation changes by at most $36\sigma=1.44$.

  Since the difference between the optimal triangulation and the best
  triangulation without the edges $xy$ and $xz$ is bigger than
  $3.83677>2\cdot 1.44$, the claim of the lemma is not invalidated if
  each point is moved by at most~$0.04$ in any direction. We will use this
  observation in Section~\ref{sec:elem-piece}, where we describe a version of 
  the wire-piece in which the offset between the terminal triangles is 
  slightly larger than in the standard wire-piece. This makes it possible
  to build wires of arbitrary length (see Lemma~\ref{long-distance}).

\subsection{Putting pieces together}
\label{sec:red-weight}

When putting pieces together, we make sure that
Lemma~\ref{lemma-two-edges} can be applied at the junction:
at every terminal triangle, the piece contains a copy of half of the
point set $W$ (possibly rotated, and possibly scaled by~$4.3$).
  This
allows us to split the input set into components at the terminal
triangles. For each terminal triangle $xyz$, we need to consider just
the two choices $xy$ and $yz$ for the \emph{terminal edges}.
We denote these choices by L and R, depending on whether we use
the left or right arms, as viewed from the tip of the terminal triangle.

 When determining the behavior of a piece with $k$ terminals in a 
minimum-weight triangulation, we have to compare the optimal
triangulations for the $2^k$ choices for the positions of the $k$
terminal edges.
However, these triangulations cover different areas and cannot be
compared directly.  Depending on whether the terminal edge $xy$ or the
terminal edge $xz$ is chosen at some given boundary, the triangulated
area excludes or includes the area of the terminal triangle itself,
and the optimal triangulation can thus be expected to be cheaper or
more expensive. To offset this effect, we define the \emph{reduced
weight} of a triangulation $T$, $\bar c(T)$, by adding a penalty or 
subtracting a bonus term from the actual weight.  This will allow us 
to see directly which configurations of terminal edges are better 
than others, and which configurations cannot possibly be part of an 
optimal triangulation.

\begin{figure}[t]
  \centering
  \includegraphics[scale=0.9]{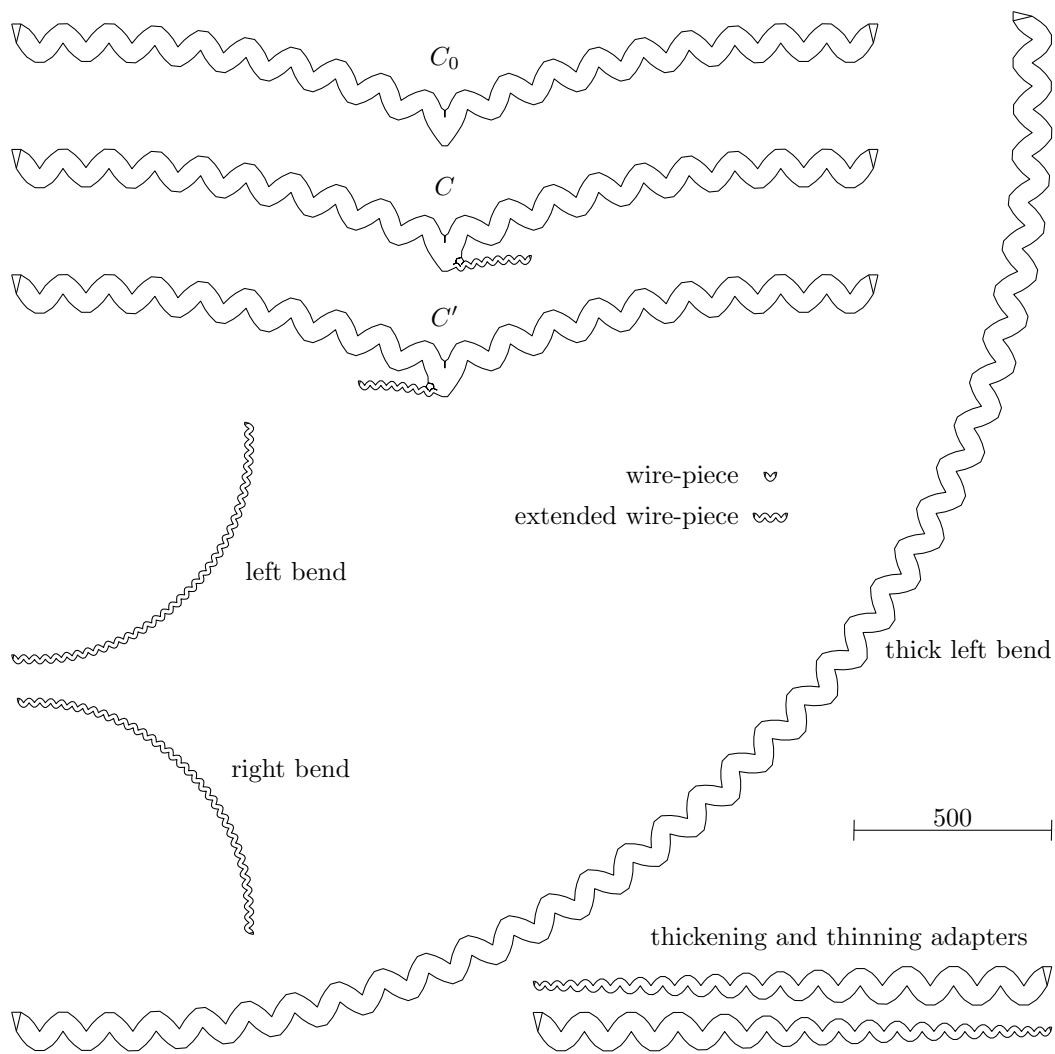}
  \caption{All elementary pieces}
  \label{all-pieces}
\end{figure}

More precisely, 
we define a difference term $\delta$ that depends on the size of the triangle
(large or small), see Table~\ref{tab:terminal}.
When comparing the costs of different triangulations of a piece or of 
different pieces, we only look at the
\emph{internal cost}, the cost of the internal edges, since the boundary
edges are fixed. When the triangulated area of a piece includes the
terminal triangle, we subtract $\delta$ from the actual weight;
otherwise we add $\delta$. The precise values of $\dsmall$ and $\dlarge$
are not important and have been chosen to make the proofs convenient. The 
following lemma holds irrespective of the values of $\dsmall$ and $\dlarge$.

\begin{lemma}\label{reduced-cost}
  Consider a layout of pieces $S$ where every terminal triangle is shared
  by two pieces.
%
  For each terminal triangle, fix one of its two long sides.  Consider
  the minimum-weight triangulation $T$ of $S$ that is restricted to
  contain those fixed edges as well as all piece boundaries.  For each
  piece, calculate the reduced internal weight of the minimum-weight
  triangulation of the piece bounded by the given fixed edges.

  Then the overall reduced internal weight of all pieces differs from
  the total weight of the triangulation $T$ by a constant that is
  independent of the choice of the fixed edges.
\end{lemma}
\begin{proof}
  In total, the effect of adding and subtracting $\delta$ cancels on
  the two sides of each terminal triangle. The cost of the terminal
  edge ($xy$ or $xz$) itself is not accounted for in the sum of
  internal costs, but since $xy$ and $xz$ have the same length, this
  amounts to a constant difference.  The boundaries are also fixed,
  and so is the cost of triangulating the pockets and the holes.
\end{proof}
We will see below (Proposition~\ref{boundary}) that all boundary edges
are part of the MWT, since they belong to the $\beta$-skeleton.
We know from Lemma~\ref{lemma-two-edges} that, by choosing a terminal
edge from each terminal triangle in all possible ways, we are
guaranteed to find the optimal triangulation.
The consequence of Lemma~\ref{reduced-cost} is that we need only look at the
reduced (internal) cost when comparing these choices.

\section{Elementary Pieces}
\label{sec:elem-piece}

\begin{figure}
  \centering
  \includegraphics[scale=0.7]{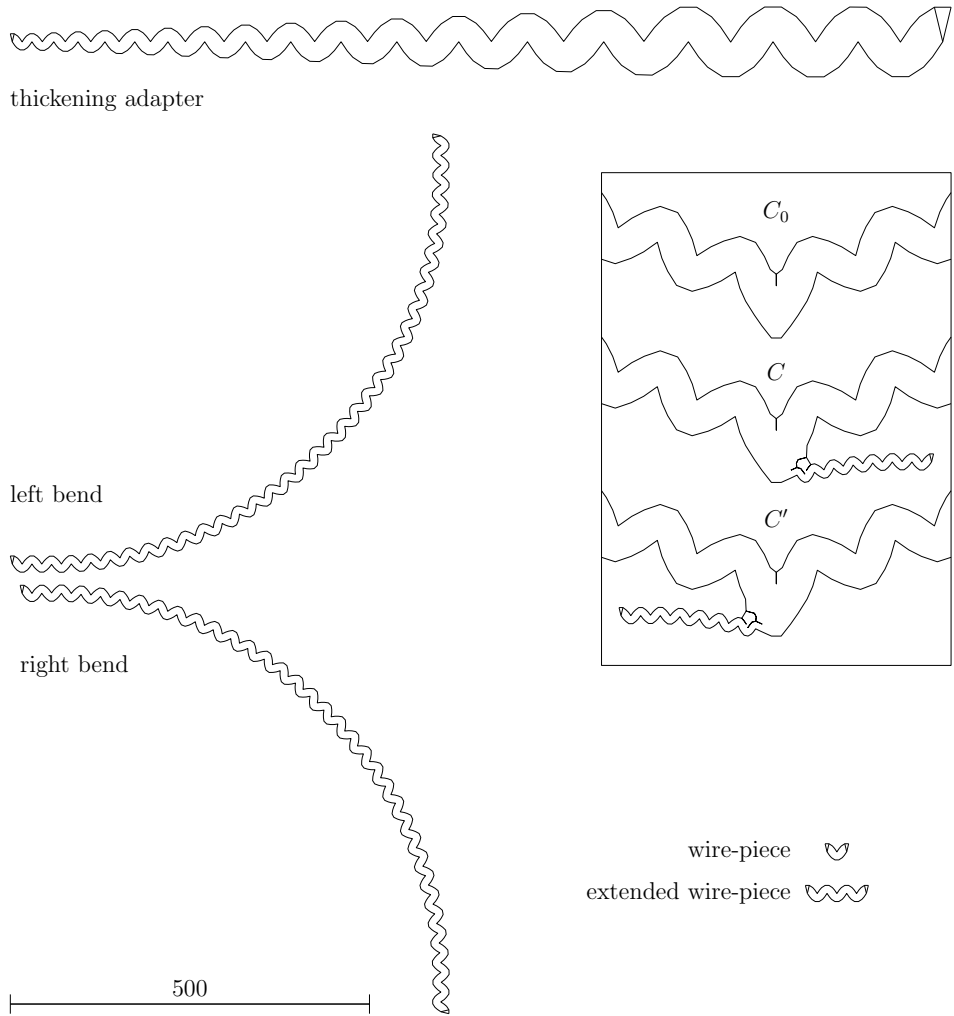}
  \caption{Enlarged view of the small elementary pieces}
  \label{small-pieces}
\end{figure}

\begin{figure}
  \centering
  \includegraphics[scale=0.9]{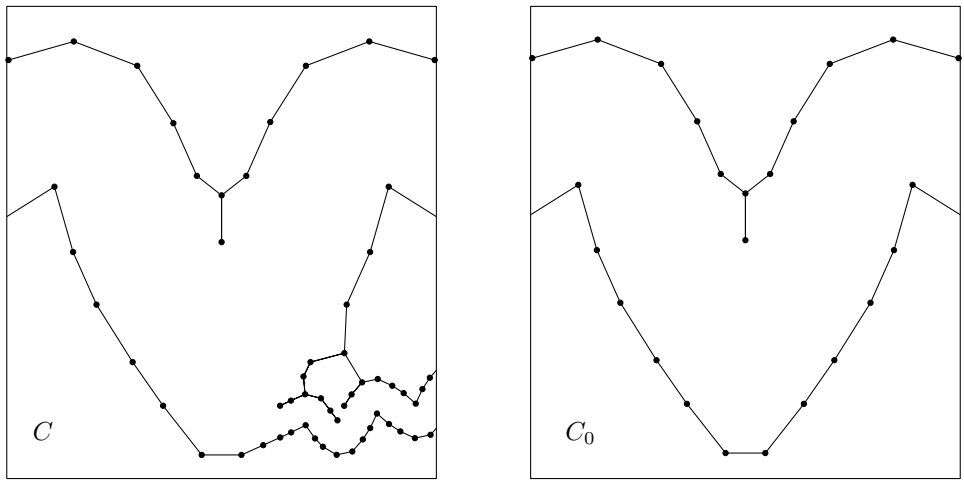}
  \caption{Enlarged view of the central part of the $C$ and $C_0$ connections.}
  \label{C-connection}
\end{figure}

\begin{figure}
  \centering
\includegraphics{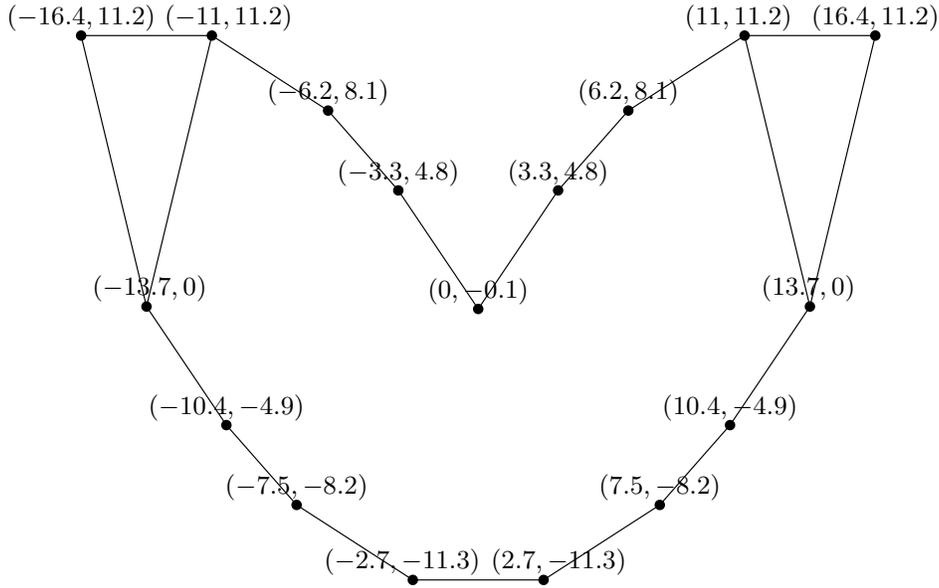}
  \caption{The coordinates of a wire-piece.}
  \label{definition-wire}
\end{figure}

We have 10 elementary pieces, shown in
Figures~\ref{all-pieces}--\ref{C-connection}: 
The (i) \emph{wire-piece}
(Figure~\ref{definition-wire})
 is the main tool achieving a non-locality for conveying information
over long distance.
As mentioned in the introduction, such wire-pieces
were originally conceived by Jack Snoeyink \cite{BeiroutiSn98}.
In order to construct wires of arbitrary length, we also have
 (ii) an \emph{extended wire-piece}.
(Figure~\ref{definition-wire}).
The horizontal offset between the two terminal triangles of a {wire-piece} is $27.4$, i.\,e., with these 
wire-pieces alone, one can bridge distances which are multiples of
 $27.4$.
The offset for the {extended wire-piece} is
$82.21=3\cdot 27.4 + 0.01$. Thus, with the right combination of
wire-pieces and extended wire-pieces, one can form a straight connection of
arbitrary length,
provided it is long enough:
\begin{lemma}\label{long-distance}
  Consider two small vertical terminal triangles at the same height
  with a horizontal distance $d>230\,000$ that is a multiple of
  $0.01$.  Then the two triangles can be connected by a sequence of
  wire-pieces and extended wire-pieces.

  An analogous statement holds for vertical connections.
\end{lemma}
\begin{proof}
  If the distance between two terminal triangles is $d=0.01 \cdot z$
  for some integer $z \geq 3 \cdot 2740 \cdot 2739 = 22\,514\,580$, they can be
  connected by concatenating $y \eqdef z \bmod 2740$ extended
  wire-pieces with $\lfloor z/2740 \rfloor - 3y$ wire-pieces.
\end{proof}

The work-horse of our gadgets is
(iii) the 
\emph{$C$-connection} and its mirror image,
(iv) the 
\emph{$C'$-connection}.
These are the only pieces with three terminal triangles,
two large ones and a small one. They serve two
purposes: they allow us to
introduce branches into the network of wires, and they effect
``negation'' of the information that is transmitted through the wires.
We also have
a (v)
\emph{$C_0$-connection} with only two terminals. It serves as a
placeholder for the $C$ or $C'$ connection when the third terminal is
not needed.

\begin{figure}
  \centering
\includegraphics{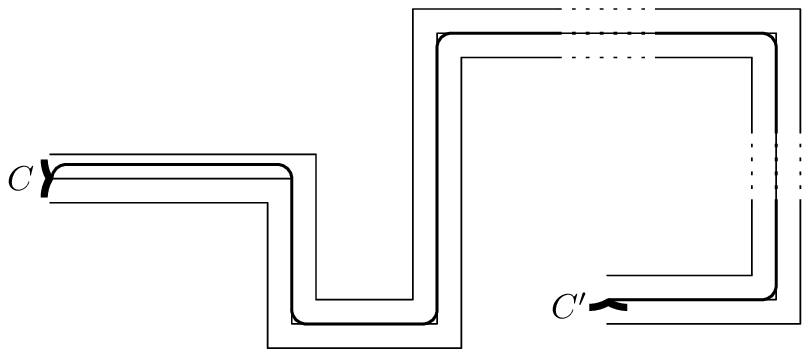}
\caption{Tracing a wire through a rectangular corridor,
forming a so-called $C$-$C'$-link (see Section~\ref{link}.)
  The dotted parts are long enough to allow connections of any length
  (Lemma~\ref{long-distance}).}
  \label{corridor-figure}
\end{figure}

The remaining pieces are variations of the wire-piece for building
more flexible wire shapes:
The
 (vi) \emph{left bend} and the
 (vii) \emph{right bend}
allow us to introduce arbitrary turns into wires. The vertical
part of the left bend corresponds to a wire-piece that is rotated $90^\circ$ 
counter-clockwise, whereas the vertical part of a right bend corresponds to 
a wire-piece that is rotated $90^\circ$ clockwise. We need both a left
and a right bend, because each piece fits to a given terminal triangle
of another piece only at one end.

With these pieces, we can arrange wires to follow any rectangular
layout, provided we blow the layout up sufficiently, see Figure~\ref{corridor-figure}:
\begin{lemma} \label{lemma-connects} Given a $C$-connection piece and
  a $C'$-connection piece and a rectangular path between them, the
  small triangles of the two pieces can be connected by a sequence of
  wire-pieces, extended wire pieces, left and right bends, within the
  corridor of width $2000$ around the path, provided that the path
  contains a straight portion of length at least $250\,000$, both in
  the horizontal and in the vertical direction, and the corridor does
  not interfere with other pieces or intersect itself.
\end{lemma}

\begin{proof}
  The left and right bend both fit into a box of size $700 \times
  700$, and by Lemma~\ref{long-distance}, we can bridge any distance
  larger than $230\,000$ in the horizontal and vertical direction. The
  wires can be positioned sufficiently far away from the boundary of
  the rectangular strip to make sure that the boundaries of different
  connections do not interfere with each other.
\end{proof}

We also have (viii) a \emph{thick left bend}, which is just a scaled
copy of the left bend.

Finally, we have two \emph{resizing wire-pieces}: the (ix)
\emph{thickening adapter} and its mirror image, the (x) \emph{thinning
  adapter}. They are used to interpolate between small and large
terminal triangles in the clause gadgets.

\iffull
\begin{figure}
  \centering
\vspace {-2,1cm} 
\includegraphics[angle=78,width=\textwidth]{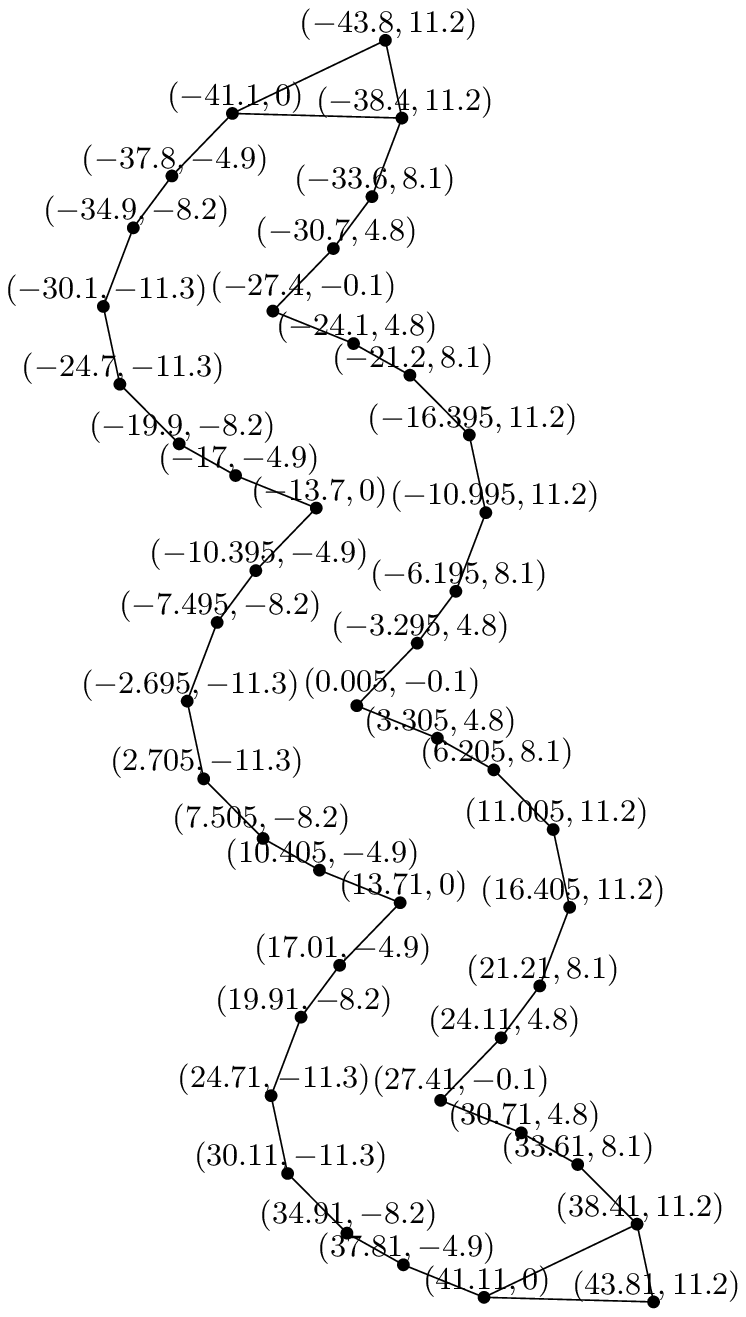}
\vspace {-2,1cm}
  \caption{The coordinates of the extended wire-piece.}
  \label{definition-wire-extended}
\end{figure}
\fi

\iffull
Figures~\ref{definition-wire} and~\ref{definition-wire-extended} show
the coordinates of the wire-piece and the extended wire-piece.
\fi
The complete data for the pieces are available
on the Internet.\footnote{%
  \url{http://www.inf.fu-berlin.de/inst/ag-ti/people/rote/Software/MWT/}%
all-coordinates.txt
  }

\begin{proposition}\label{basic-properties}
The elementary pieces have the following properties:

All coordinates are multiples of $0.0001$.  Each piece contains two or
three copies of a terminal triangle, together with a copy of $\Wleft$
or $\Wright$ \textup(Figure~\ref{set-W}\textup),
possibly rotated by $90^\circ$, and possibly scaled
by~$4.3$\textup.  The coordinates of the terminal triangles are multiples of
$0.01$.
\qed
\end{proposition}
Each piece has two terminal triangles, with the exception of the
connection pieces, which have three terminal triangles, as shown in
the figures.
The edges shown in the figure, with the exception of
 the two equal sides of each terminal triangle,
form the \emph{boundary} of the pieces.
\begin{proposition}\label{boundary}
  The boundary edges of all elementary pieces are contained in the
  $\beta$-skeleton, for $\beta=1.1806$. This remains true when the pieces
  are connected using their terminal triangles, as long as the
  boundary part of different pieces are sufficiently far apart.
  Hence, all boundary edges belong to the MWT.
\end{proposition}

\begin{proof}
The lemma can be verified manually by inspecting the relevant 
regions for the boundary edges. Since the number of boundary edges is very 
large, we have implemented a computer program which computes the 
$\beta$-skeleton. 
\iffull
The source code can be found in 
Appendix~\ref{app:beta-skeleton}.
\fi
\end{proof}

The boundary decomposes 
into two or three connected
\emph{boundary} pieces.  As can be seen in Figure~\ref{C-connection},
some boundary pieces are not just paths, but they form trees.


\subsection{Analysis of the Pieces}
\label{piece-analysis}

In this section we discuss the exact properties of our gadgets as they were
computed by our computer program. We present tables which show the internal
cost, the reduced internal cost $\bar c$ (see Section~\ref{sec:red-weight}) and the
\emph{relative reduced internal cost $\tilde c$} (i.e., the difference between the reduced cost
of a certain configuration and the minimum reduced cost incurred by any 
configuration of a gadget).

\begin{figure}[htb]
  \centering
\hbox to \textwidth{\hfill
\includegraphics[scale=0.77] {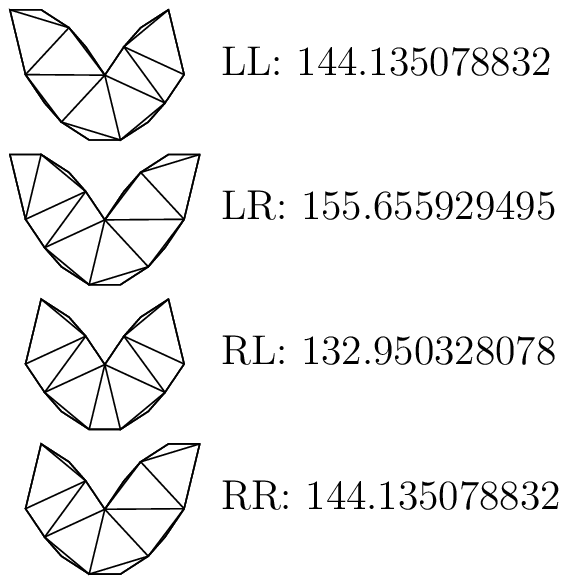}\hfill\hfill
\includegraphics[scale=0.77] {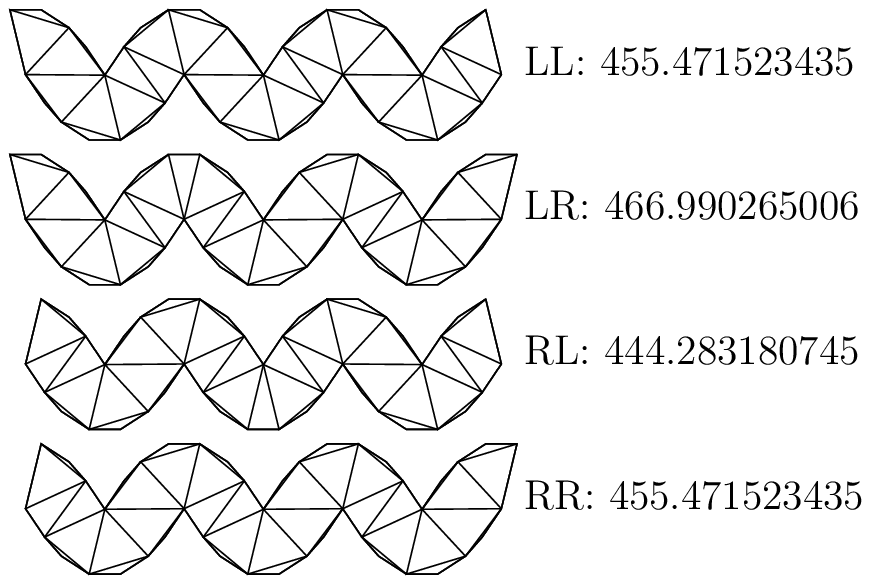}\hfill}
  \caption{Optimal solutions for all cases for the wire-piece 
and the extended wire-piece}
  \label{fig:TABLE-wire}
%
  \label{fig:TABLE-wire-extended}
\end{figure}

For example, let us look at the extended wire-piece.  It has two
terminal triangles.  For each of the four combinations of states, LL,
LR, RL, and RR, the right part of Figure~\ref{fig:TABLE-wire-extended} shows 
an optimal solution, together with the internal cost.
Table~\ref{tab:extended-piece} summarizes this information. The second column 
lists the number of optimal solutions (or potential optimal solutions, i.\,e.,
all solutions which the program could not exclude from being an optimal 
solution, with the given precision of the calculation). Indeed, looking
back at Figure~\ref{fig:TABLE-wire-extended}, one can see 
that the solution for case LR has a symmetric copy with the same cost. 
The astute reader may wonder why the multiplicities for cases LR and RL are
not larger. For example, it looks as if the isosceles ``$xyz$'' triangle for 
case LR could be placed also at the very left end or at the very right end.
However, this is not the case, since the three repetitions of the wire-piece 
that constitute the extended wire-piece are not exact repetitions:
the ends have been moved apart, stretching the middle piece
and thus causing a deviation from the regular periodical pattern.
The third column of Table~\ref{tab:extended-piece} lists the internal cost, 
repeating the information from Figure~\ref{fig:TABLE-wire-extended}.

\begin{table}[htb]
  \centering
\noindent\vbox{  
    \offinterlineskip
    \halign{\vrule \strut\hfil\ #\hfil \vrule &\ \hfil $#$ \hfil\vrule
    &&\ \hfil $#$ \hfil\vrule \cr
     \noalign{\hrule}
     \raise2pt\hbox{\strut}%
     pattern & $multiplicity$ &$internal cost $ c& $reduced internal
     cost $ \bar c
     &$relative reduced cost $ \tilde c\cr \noalign{\hrule}
    \noalign{\hrule}LL &1& 455.471\,523\,435 & 455.471\,523\,435 & 0.000\,000\,000 \cr
LR &2& 466.990\,265\,006 & 455.679\,921\,006 & 0.208\,397\,570 \cr
RL &1& 444.283\,180\,745 & 455.593\,524\,745 & 0.122\,001\,310 \cr
RR &1& 455.471\,523\,435 & 455.471\,523\,435 & 0.000\,000\,000 \cr
\noalign{\hrule}
}}
  \caption{Analysis of the extended wire-piece}
  \label{tab:extended-piece}
\end{table}

The fourth column is the reduced internal cost $\hat c$, obtained by adding or
subtracting $\dsmall$, as appropriate. For example, for case LR, we
have to subtract $2\dsmall = 11.310\,344$, since this triangulation
covers the larger area on both the left and the right end.
For cases LL and RR, addition and subtraction of $\dsmall$
cancel, and the reduced cost remains unchanged.

The rightmost column reduces all numbers by subtracting the column
minimum, to make the differences more visible.  It is now plain to see
that LL and RR have the same cost, which is also obvious by symmetry.
We denote these \emph{relative reduced costs} by $\tilde c$, and we
think of them as \emph{penalties} for deviating from the ``ground
state'' where the smallest reduced cost is achieved.  They are
non-negative by definition.

\iffull
The corresponding table for all pieces, together with pictures of the optimal 
triangulations, is given in Table~\ref{tab:pieces} of 
Appendix~\ref{appendix-pieces}.
\fi
Table~\ref{tab:pieces-summary} condenses the information for all
pieces with two terminals.

Since the pieces are symmetric, LL and RR always have the same cost.
The only exceptions are the resizing pieces.
%
Let $\eps_1 \approx 0.000\,051$ be the relative reduced cost of  the
thickening adapter in state LL.

\begin{table}[htb]
  \centering
\noindent \vbox{\hbox
   {\valign{\hrule\hbox {\vrule\ \raise1pt\hbox{\strut}\strut # }\hrule
       &\hbox {\vrule\ \raise2pt\hbox{\strut}\strut\ # }%
       &&\hbox {\vrule\ \strut\ # }\cr
       & LL& LR& RL& RR\cr
wire-piece& 0.000\,000\,000& 0.210\,506\,663& 0.125\,593\,246& 0.000\,000\,000\cr
extended wire-piece& 0.000\,000\,000& 0.208\,397\,570& 0.122\,001\,310& 0.000\,000\,000\cr
thickening adapter& $0.000\,051\,402\ldots=\eps_1$& 0.018\,887\,246& 0.014\,627\,250& 0.000\,000\,000\cr
thinning adapter& 0.000\,000\,000&0.018\,887\,246& 0.014\,627\,250& $0.000\,051\,402\ldots=\eps_1$\cr
}\vrule}\hrule}

\medskip

\noindent \vbox{\hbox
   {\valign{\hrule\hbox {\vrule\ \raise1pt\hbox{\strut}\strut # }\hrule
       &\hbox {\vrule\ \raise2pt\hbox{\strut}\strut\ # }%
       &&\hbox {\vrule\ \strut\ # }\cr
       & LL& LR& RL& RR\cr
$C_0$ connection& 0.000\,000\,000& 0.044\,001\,701& 0.020\,571\,757& 0.000\,000\,000\cr
 left bend& 0.000\,000\,000& 0.086\,460\,895& 0.125\,593\,246& 0.000\,000\,000\cr
 right bend& 0.000\,000\,000& 0.210\,506\,663& 0.125\,593\,246& 0.000\,000\,000\cr
thick left bend& 0.000\,000\,000& 0.891\,261\,046& 0.020\,571\,757& 0.000\,000\,000\cr
}\vrule}\hrule}

  \caption{The relative reduced costs $\tilde c$ for the pieces with two terminals}
  \label{tab:pieces-summary}
\end{table}

The $C$-connection (and its mirror image, $C'$) is the only piece with
three terminals. The results for this piece is shown in
\iffull
Table~\ref{tab:pieces-C}, and in more visual form in
Figure~\ref{fig:pieces-C}.
\fi
 We encode the states by three letters $ABc$,
where the capital letters $A$ and $B$ refer to the left and right large
 terminal triangles (L or R), as usual, and $c$ refers to the
small terminal triangle (l or r).

\iffull
\begin{table}[htb]
  \centering
\noindent\vbox
{
    \offinterlineskip
    \halign{\vrule \strut\hfil\ #\hfil \vrule &\ \hfil $#$ \hfil\vrule
    &\ \hfil $#$ \hfil\vrule
    &\  $#$ \hfil\vrule
    &\ \hfil $#$ \hfil\vrule
 \cr
     \noalign{\hrule}
     \raise2pt\hbox{\strut}%
     $C$ &$internal cost $c & $reduced internal cost $\hat c
     &\hfil$relative reduced cost $\tilde c&C'\cr \noalign{\hrule}
     \noalign{\hrule}
LLl &14\,027.752\,986\,494 & 14\,033.408\,158\,494 &0.003\,861\,076\ldots
=\delta_1 &$RRr$\cr
LLr &14\,039.059\,470\,993 & 14\,033.404\,298\,993 &0.000\,001\,575\ldots
=\eps_2 &$RRl$\cr
LRl &14\,075.894\,120\,134 & 14\,033.434\,292\,134 &0.029\,994\,716 &$LRr$\cr
LRr &14\,087.200\,604\,634 & 14\,033.430\,432\,634 & 0.026\,135\,215&$LRl$\cr
RLl &13\,979.654\,697\,176 & 14\,033.424\,869\,176 & 0.020\,571\,757&$RLr$\cr
RLr &13\,990.965\,042\,750 & 14\,033.424\,870\,750 & 0.020\,573\,332&$RLl$\cr
RRl &14\,027.749\,125\,419 & 14\,033.404\,297\,419 & 0.000\,000\,000=0 &$LLr$\cr
RRr &14\,039.064\,100\,296 & 14\,033.408\,928\,296 & 0.004\,630\,878\ldots
=\delta_2 &$LLl$\cr
\noalign{\hrule}
}
}
  \caption{Analysis of the pieces $C$ and $C'$. Since $C'$ is the
    mirror image of $C$, it is represented in the same table}
  \label{tab:pieces-C}
\end{table}
\fi

{\let\orignormalsize\normalsize
\iffull
\begin{figure}
\else
\begin{table}
\fi
 \centering
\noindent
\hbox to 0pt
{\hss\includegraphics[scale=0.9]{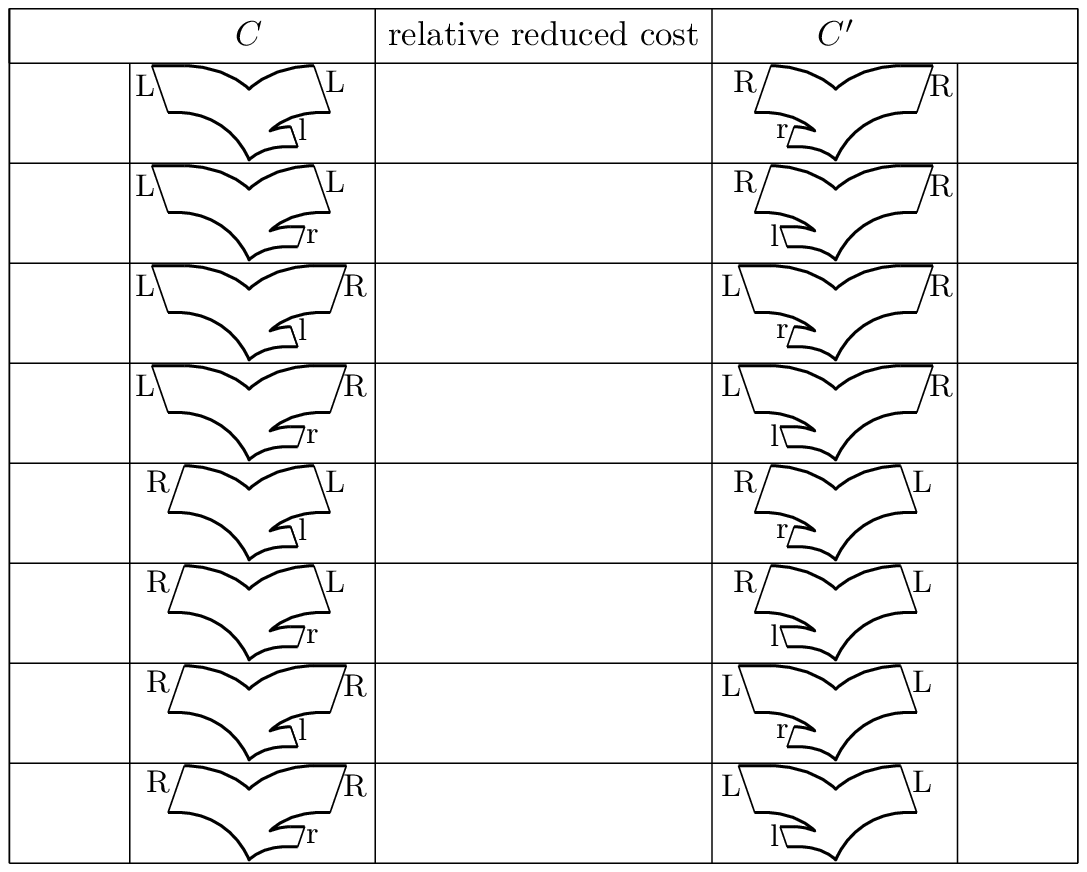}\hss}%
\hbox to 0pt
{\hss \vbox{
\baselineskip=0,9156cm
\halign{\orignormalsize#\hskip 25,5mm\hfil&$#$\hfil\hskip 26mm&\orignormalsize#\hfil\cr
LLl  &0.003\,861\,076\ldots=\delta_1 &RRr\cr
LLr  &0.000\,001\,575\ldots=\eps_2 &RRl\cr
LRl  &0.029\,994\,716 &LRr\cr
LRr  & 0.026\,135\,215&LRl\cr
RLl  & 0.020\,571\,757&RLr\cr
RLr  & 0.020\,573\,332&RLl\cr
RRl  & 0.000\,000\,000=0 &LLr\cr
RRr  & 0.004\,630\,878\ldots=\delta_2 &LLl\cr
}\vskip 3,5mm}\hss}
  \caption{All cases for the
analysis of the pieces $C$ and $C'$.
}
\iffull
  \label{fig:pieces-C}
\end{figure}
\else
  \label{tab:pieces-C}
\end{table}
\fi
}

All tables were computed using fixed-precision interval arithmetic
with 15 decimal digits after the decimal point (i.\,e., multiples of
$10^{-15}$).
The resulting intervals were rounded to 9 digits for displaying. It turned out that the intervals
were small enough so that they could be rounded to unique 9-digit
numbers.
Thus, for example,
$0.000\,051\,402$
 denotes a number that is guaranteed to lie strictly
between
$0.000\,051\,401\,5$ and
$0.000\,051\,402\,5$.

By inspecting Tables~\ref{tab:pieces-summary} and~\ref{tab:pieces-C},
one can see that there is a penalty for having two triangles of
different states.  If both terminal triangles of a piece with two
terminal triangles, or the two large terminal triangles of $C$ or $C'$,
are in a different state, we call this a \emph{breach}.
A breach is so expensive that it will never occur in an optimal
solution.
We summarize this information for later use.
\begin{proposition}\label{prop:equal-large}
  In every piece, a breach has relative reduced cost at least
$\delta_3 \eqdef 0.01$.
If there is no breach, the
 relative reduced cost is zero, except
for $C$, and $C'$, and the resizing wire-pieces.
For the resizing wire-pieces, the relative reduced cost is
at most $\eps_1 < 0.000{\,}051$.
%
%
%
\qed
\end{proposition}

It follows that long wires have a preference to be triangulated 
uniformly:

\begin{lemma}[(The Wire Lemma)]\label{lemma:uniform-wire}
  Let $T_1$, $T_2$ be two small terminal triangles connected by an
  arbitrary sequence of wire-pieces, extended wire-pieces, 
  and left and right bends. If $T_1$ and $T_2$ are in the 
  same state, all the connecting pieces are in that state and the 
  relative reduced cost is 0. If $T_1$ and $T_2$ are in different
  states,
  the relative reduced cost is at least $\delta_3=0.01$.
  \qed
\end{lemma}

The states of $C$ and $C'$ without a breach are LLl, LLr, RRl, and
RRr.  {From} Table~\ref{tab:pieces-C}, one sees that the preferred states are LLr,
and RRl, where the small triangle has the opposite state from the
large triangles.  We call these two states the \emph{consistent}
states, and the states LLl and RRr \emph{inconsistent}.
We define
\begin{align*}
 \eps_2 &\eqdef 
\tilde c(C,\mathrm{LLr}) =
\bar c(C,\mathrm{LLr}) - \bar c(C,\mathrm{RRl})
=
 c(C,\mathrm{LLr}) -  c(C,\mathrm{RRl}) - 2\dsmall
\\\delta_1 &\eqdef 
\tilde c(C,\mathrm{LLl}) =
\bar c(C,\mathrm{LLl}) - \bar c(C,\mathrm{RRl})
=
 c(C,\mathrm{LLl}) -  c(C,\mathrm{RRl})
\\\delta_2 &\eqdef 
\tilde c(C,\mathrm{RRr}) =
\bar c(C,\mathrm{RRr}) - \bar c(C,\mathrm{RRl})
=
 c(C,\mathrm{RRr}) -  c(C,\mathrm{RRl}) - 2\dsmall,
\end{align*}
where $c(C, \mathrm{LLr})$, $\bar c(C, \mathrm{LLr})$, etc.\ denote
the 
cost of the $C$-connection in the
respective configuration, see Table~\ref{tab:pieces-C}.
\begin{proposition}\label{prop:C-consistent}
In $C$ or $C'$,
every consistent state has
 relative reduced cost at most
$\eps_2 < {0.000\,002}$.
The two inconsistent states have cost
$\delta_1  \approx 0.003{\,}861$ and
$\delta_2  \approx 0.004{\,}631$.
Thus, if there is no breach in $C$ or $C'$, the
 relative reduced cost is at most
$\delta_2$.
\qed
\end{proposition}

In general, we denote by $\eps_1$, $\eps_2$, ``small'' quantities that
we would like to neglect, and which can be made arbitrarily small by
refining the coordinates of the pieces or the digits of $\dsmall$ and
$\dlarge$.  On the other hand, $\delta_1$, $\delta_2$, \ldots,
denote ``large'' quantities whose difference will be made productive
for the proof.

\section{Larger Gadgets}

We now describe how to assemble the elementary pieces from the last section
to obtain larger gadgets which model the logical structure of a 
PLANAR 1-IN-3-SAT formula. We need \emph{variables}, \emph{clauses}, and the
connections between them. The main building block for the variables and
clauses is the \emph{bit loop} (Section~\ref{sec:vloop}). Each bit
loop stores a logical state (L or R). To ensure consistency between different 
bit loops, we use \emph{$C$-$C'$-links} (Section~\ref{link}). By 
connecting several bit loops, we build clauses 
(Section~\ref{sec:clause-gadget}) and variables (Section~\ref{sec:var}).
The optimal triangulation of a clause is obtained if and only if exactly one
of its inputs is in state L (the 1-IN-3 property). A variable is
modeled as a 
chain of bit loops connected by $C$-$C'$-links so that all loops are in
the same state. Variables can be connected to clauses using the
wire-pieces from the last section.

\subsection{The Bit Loop}
\label{sec:vloop}
\begin{figure}
  \centering
  \includegraphics[scale=0.7]{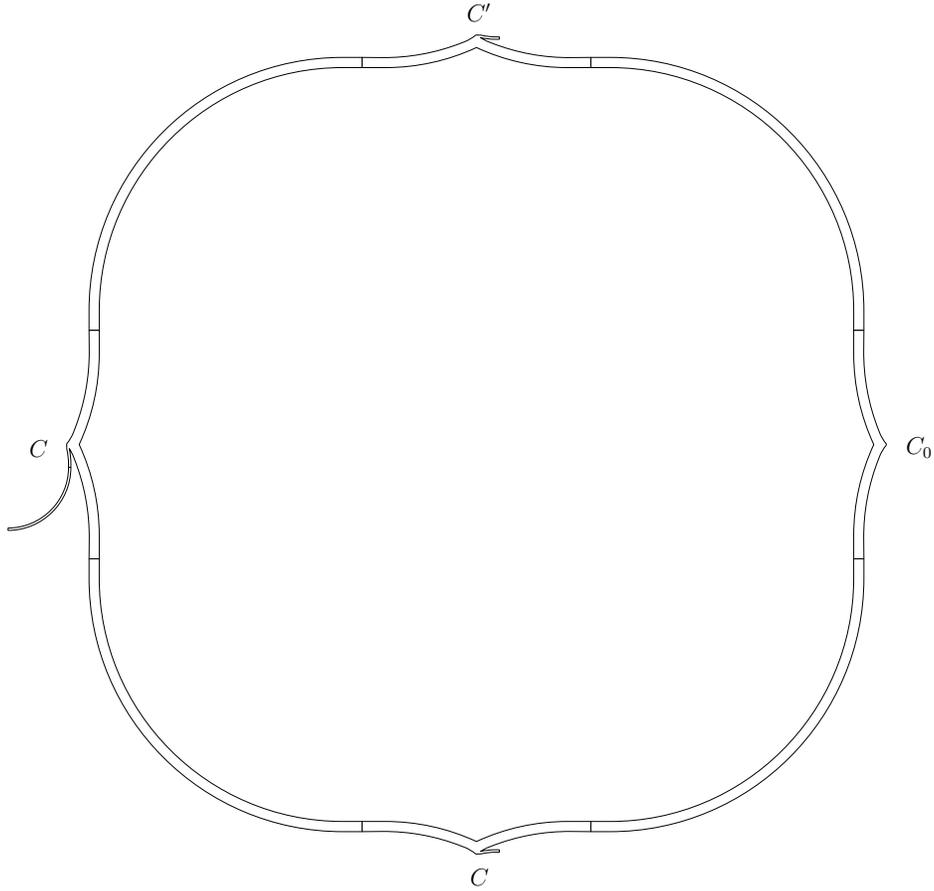}
  \caption{An instance of a bit loop with two $C$-connections and
    one $C'$-connection. At the left $C$-connection, a right bend is
    attached, as the beginning of a wire joining it to another
    $C'$-connection. The drawing is to scale, but the shapes are
    simplified,}
  \label{variable-loop}
\end{figure}

The \emph{bit loop} is used to represent a logical state (L or R).
It if formed by connecting any selection of four connection 
pieces ($C$, $C'$
or $C_0$) into a loop, using four thick bends, as shown schematically
in Figure~\ref{variable-loop}.

\begin{lemma} \label{variable-equal}
In an optimal triangulation, the 8 large terminal triangles of a
bit loop are in the same state \textup(L or R\textup).
\end{lemma}
\begin{proof}
  The only interaction with other pieces is through the at most four
  small ``exit'' triangles.
  Let us keep the states of these triangles
  fixed.  
  If we set all 8 large terminal triangles to L or to R, the
  maximum relative reduced cost is at most $4\cdot\delta_2 <
  0.02=2\delta_3$, see
Proposition~\ref{prop:C-consistent}. 

  On the other hand, if the 8 large terminal
  triangles are not all equal, there are at least two breaches.
Hence, by Proposition~\ref{prop:equal-large},
  the relative reduced cost is at least $2\delta_3$, and such a
  triangulation cannot be optimal.
\end{proof}

We can therefore simply refer to the \emph{state of the bit loop}
as the common state of its large terminal triangles.  Generally, when
it is clear that there is no breach, the state of a $C$, $C'$ or
$C_0$-connection refers to the state of its large terminal triangles.

\subsection{The $C$-$C'$-Link}
\label{link}

A $C$-connection can be joined to a $C'$-connection by an arbitrary
sequence of wire-pieces,
extended wire-pieces, and left and right bends
(Lemma~\ref{lemma-connects}),
see Figure~\ref{corridor-figure}.
We call the two connections at the end together with their joining wire
a \emph{$C$-$C'$-link}.

$C$-$C'$-links are used in two contexts:
The most common case is when the $C$-connection of the $C$-$C'$-link
belongs to a bit loop $V_1$ and the $C'$-connection belongs to
another bit loop $V_2$. In the clause gadget 
(Section~\ref{sec:clause-gadget}), we also have $C$-$C'$-links where
the $C$-connection is connected to a thickening adapter and a thinning adapter,
while the $C'$ connection is part of a bit loop.
By Lemma~\ref{variable-equal} we know that in each of $V_1$ and $V_2$,
the large triangles are in a uniform state (L or R), and in 
Section~\ref{sec:clause-gadget} we will see that an analogous statement
holds for the large terminal triangles of the thickening and the thinning
adapter. We thus have to
consider four cases for these states, as shown in
Table~\ref{tab:C-connection}.  For each combination, we can work out
the optimum state of the small terminal triangles by considering
Table~\ref{tab:pieces-C}. (The first two and last two rows of
Table~\ref{tab:pieces-C} are sufficient.)
The optimal choice for the thin wire is shown in the third column.

\begin{table}[htb]
  \centering
  \begin{tabular}{|c|c|c|c|c|l|}
\hline
    $V_1$ & $V_2$ & wire
 & state of $C$
 & state of $C'$
&optimum relative reduced cost\\
\hline
L & L & r & LLr & LLr & $\tilde c(C,\mathrm{LLr})+\tilde c(C',\mathrm{LLr})=\eps_2$
\\
L & R & l (or r) & LLl & RRl &
$\tilde c(C,\mathrm{LLr})+\tilde c(C',\mathrm{RRr})=\eps_2+\delta_1$
\\
R & L & l (or r)& RRl & LLl &
$\tilde c(C,\mathrm{RRr})+\tilde c(C',\mathrm{LLl})=\delta_2$
\\
R & R & l & RRl & RRl &
$\tilde c(C,\mathrm{RRl})+\tilde c(C',\mathrm{RRl})=\eps_2$
\\
\hline
  \end{tabular}
  \caption{The best state of the wire for a $C$-$C'$-link.
For each given combination of states (first two columns), the last
column gives the resulting cost.}
  \label{tab:C-connection}
\end{table}

\begin{lemma}\label{lem:C-connection}
  Each $C$-$C'$-link incurs a relative reduced cost of
at least $\eps_2$.
The cost reaches this minimum if and only if the large terminal
triangles in the two pieces have the same state. Otherwise,
the cost increases to $\delta_1+\eps_2$ or $\delta_2$, respectively,
see Table~\ref{tab:C-connection}.
\end{lemma}

\begin{proof}
  We need only exclude the possibility that the two small terminal
  triangles are in different states. By the Wire Lemma (Lemma~\ref{lemma:uniform-wire}),
  this would cause a relative reduced cost of at least 
  $\delta_3 = 0.01$, which is larger than any of the costs in 
  Table~\ref{tab:C-connection}.
\end{proof}
We call the $C$-$C'$-link \emph{inconsistent} if the large triangles
in the two connection pieces have different states.  An
inconsistent $C$-$C'$-link incurs a relative reduced cost bigger than
$\delta_1$.

\subsection{The Clause Gadget}
\label{sec:clause-gadget}

\begin{figure}[htb]
  \centering
  \includegraphics[scale=0.8]{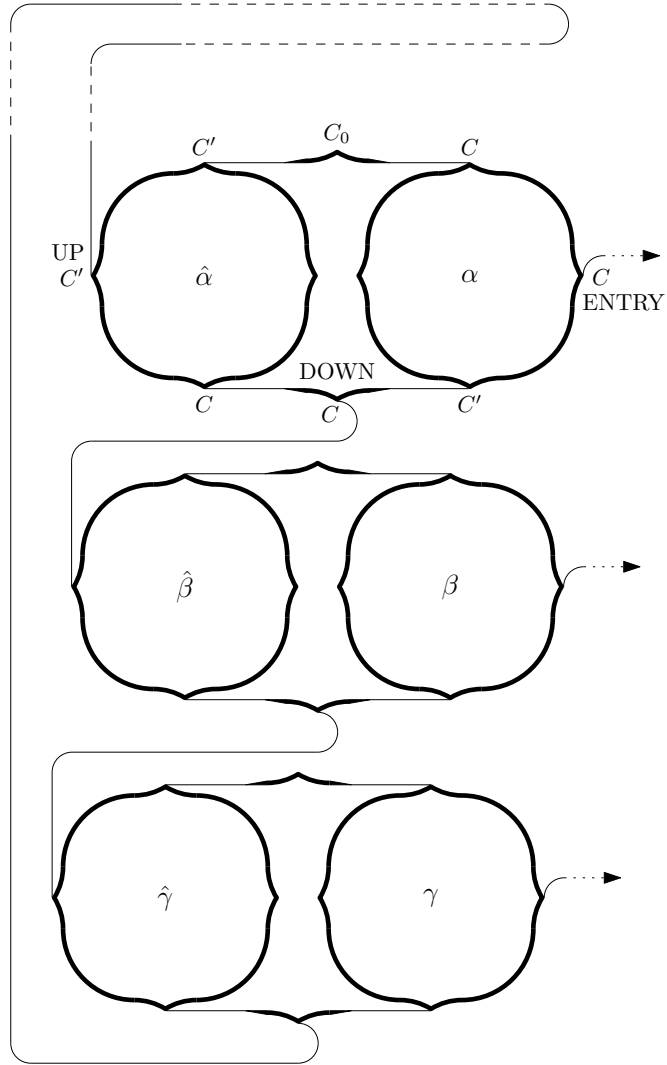}
  \caption{Schematic view of the clause gadget. The dashed parts of
    the wires are long enough for sufficiently many copies of
the wire-piece and the extended wire-piece to ensure that
the wire can reach from the lowest $C$ connection to the left $C'$
connection
of $\hat\alpha$ (Lemma~\ref{long-distance}).}
  \label{fig:clause}
\end{figure}

The \emph{clause} gadget is formed by 3 pairs of bit loops,
$\alpha$ and $\hat \alpha$, $\beta$ and $\hat \beta$, $\gamma$ and~$\hat \gamma$, see Figure~\ref{fig:clause}.  The two loops of each
pair are connected by a $C$-$C'$-link in which a thickening
adapter, a $C$-connection piece labeled DOWN, and a thinning adapter is
interspersed, and another similar $C$-$C'$-link with a
$C_0$-connection piece in the middle.
The $\alpha$-$\hat \alpha$ group has three \emph{external}
connections:
DOWN, UP, and ENTRY.

The DOWN $C$-connection piece of the pair $\alpha$-$\hat \alpha$ is
connected to the UP $C'$ connection of $\hat\beta$, and so on in a
circular way. Finally, each of $\alpha$, $\beta$, and $\gamma$, has a
$C$-connection piece, labeled ENTRY, which will be connected to one of
the ``input'' variables of the clause by a $C$-$C'$-link.

\begin{lemma}
\label{lem:clause-triangulation}
  In an optimal triangulation,
$\alpha$ and $\hat \alpha$ are in the same state \textup(L or R\textup),
and the two large terminal triangles in the DOWN $C$-connection 
are in the opposite state. 

Analogous statements hold for $\beta,\hat \beta$ and $\gamma,\hat \gamma$.
\end{lemma}
\begin{proof}
  By symmetry, it suffices to consider $\alpha$ and $\hat \alpha$.
  The $\alpha$-$\hat \alpha$ group has three external triangles UP,
  DOWN, and ENTRY which connect it to the outside world.

  There are two ``ground states'' with no internal breaches or
  inconsistencies. In these states, $\alpha$ and $\hat\alpha$ (including UP
  and ENTRY) have the same state and DOWN and the upper
  $C_0$-connection piece in the middle are in the opposite state.

If we ignore the three external connections, the relative reduced cost
of these two states is $2\eps_1 + 2\eps_2$: The contribution from the
two $C$-$C'$-links is $2\eps_2$.  The contribution from the four
resizing wire-pieces is $2\eps_1$ since we have two thickening and two
thinning adapters: the ``penalty state'' for the thickening adapter
(LL) is the optimal state for the thinning adapter, and vice versa.

Now, fix a setting for the three small external triangles.
If we select the ground state which is consistent with the majority of
the three exit triangles, we have at most one inconsistent exit
connection, causing an additional penalty of at most $\delta_2+2\eps_2$
(two consistent connections with at most $\eps_2$, and one
inconsistent connection with at most $\delta_2$).

Thus, there is always a ground state solution with cost at most
$\delta_2+2\eps_1 + 4\eps_2<0.005$.
If $\alpha$ and $\hat \alpha$ are not in the same state,
or if DOWN is in the same state as $\alpha$ and $\hat \alpha$,
 there must be
at least two inconsistencies (or even breaches) in the connections
between 
 $\alpha$, $\hat \alpha$, and DOWN,
causing a cost 
 of at least $2\delta_1  > 0.007$, which is bigger than for the ground
 state solution.
\end{proof}

Since DOWN is always in a different state from UP, ideally,
to successive loops in the cyclic sequence $\alpha\beta\gamma$ should
be in a different state. However, since there are three loops, there has
to be at least one inconsistency, which provides us the asymmetry
necessary for a 1-IN-3-clause.

Thus, we obtain:
\begin{lemma}\label{lem:clause-cost}
  The reduced cost of the clause gadget \textup(excluding the three
  ENTRY $C$-connections\textup) achieves its minimum if and only if
  exactly one of $\alpha$, $\beta$, and $\gamma$
  is in state $\mathrm{L}$. Any other triangulation incurs a cost that
  is at least $\delta_4 \eqdef 0.0007$ larger.
\end{lemma}

\begin{proof}
By Lemma~\ref{lem:clause-triangulation},
it suffices to analyze the 8 possible configurations
of the pairs $\alpha$-$\hat \alpha$, $\beta$-$\hat \beta$, and
$\gamma$-$\hat \gamma$ (ignoring the contribution of the ENTRY
connections to the clause).

The ``internal'' contribution or each pair of $2\eps_2+2\eps_1$ from
its resizing pieces and its two internal $C$-$C'$-links is constant,
and thus we can ignore this amount when comparing the various
possibilities.

Let us look at the $C$-$C'$-link between two successive pairs, say
 $\alpha$-$\hat \alpha$ and $\beta$-$\hat \beta$.
If they are equal, they cause an inconsistency, and the
relative reduced cost is
$\delta_2$ if $\alpha=\beta=\textrm{L}$
and $\delta_1+\eps_2$ if $\alpha=\beta=\textrm{R}$,
according to
 Table~\ref{tab:C-connection}. 
If $\alpha$ and $\beta$ are in different states, the link is
consistent, and the cost is $\eps_2$.

Since the situation is unchanged under cyclic shifts of
the sequence $\alpha \beta \gamma$, it is enough to consider four
cases:
\begin{itemize}
  \item $\alpha \beta \gamma = \text{LLL}$ 
%
or
 $\alpha \beta \gamma = \text{RRR}$:
In these cases, we have three inconsistencies, and 
the relative reduced cost is bigger than
    $3 \delta_1 > 0.01$.
  \item 
 $\alpha \beta \gamma = \text{LLR}$:
the relative reduced cost is $\delta_2 + 2\eps_2 > 0.0046$.
 \item 
 $\alpha \beta \gamma = \text{RRL}$:
the relative reduced cost is $\delta_1 + 3  \eps_2  < 0.0039$.
\end{itemize}
%
\end{proof}

In the previous lemma, we have ignored the relative reduced cost of the ENTRY
$C$-connections. They will be accounted for as part of the
$C$-$C'$-links that they form with the variables, to be described next.

\begin{figure}[htb]
  \centering
  \includegraphics[scale=0.8]{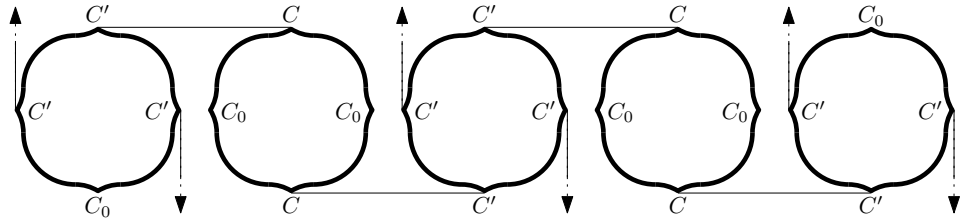}
  \caption{Schematic view of a variable gadget with three upper and three lower exits. The pattern can be repeated arbitrarily.}
  \label{fig:variable}
\end{figure}

\subsection{The Variable Gadget}
\label{sec:var}

The gadget that represents a variable of the 1-IN-3 instance is formed by 
chaining together sufficiently many
bit loops and connecting them in a row using
$C$-$C'$-links, as shown in Figure~\ref{fig:variable}.  The
dotted arrows denote potential exit wires that establish
$C$-$C'$-links with the corresponding clause gadgets
of the previous section. 
Unused exits can be replaced by
$C_0$ connections.

\begin{lemma}\label{lem:variable-triangulation}
Consider a variable chain $V$ together with the
 $C$-$C'$-links that connect it to the clauses,
including the ENTRY $C$-connections in the clauses.
The reduced cost of this point set 
achieves its minimum if and only if all the bit loops and the 
$C$-connections in the clauses are in the same state. Any other 
triangulation incurs a cost that is at least $\delta_1$ larger.
\end{lemma}

\begin{proof}
This follows directly from Lemmas~\ref{variable-equal}
and~\ref{lem:C-connection}.
\end{proof}

\section{The Reduction}
\label{sec:reduction}

\begin{theorem}\label{reduction}
Minimum-weight triangulation is strongly NP-hard.
\end{theorem}

\begin{proof}
A rectilinear embedding of the given PLANAR 1-IN-3-SAT formula can be
constructed on a grid of size $O(n)\times O(n)$.
The reduction procedure then simply replaces the edges, variables and clauses 
of the PLANAR 1-IN-3-SAT formula by the appropriate gadgets, taking care to 
leave enough space between the individual wires.

This procedure yields a point set $S$.
By construction, the boundaries of all the gadgets are part of
the $\beta$-skeleton of $S$
(Proposition~\ref{boundary}), and therefore they belong to the
minimum-weight triangulation.
The faces outside the wires are simple polygons and can be optimally triangulated using
dynamic programming. For each gadget, we know the desired
``ideal triangulation''
(Lemmas~\ref{lemma:uniform-wire} and \ref{variable-equal}--%
\ref{lem:variable-triangulation}) and can calculate its weight. Adding
up these weights and the weights of the faces outside the wires yields
a target weight $w$.  By construction, the input instance is 1-IN-3
satisfiable if and only if the minimum weight of a triangulation of
$S$ is $w$.  Otherwise, the weight of the shortest triangulation is at
least $w + 0.0007$ (Lemmas~\ref{lem:clause-cost}
and~\ref{lem:variable-triangulation}).

The set $S$ %
is a subset of an $O(n)\times O(n)$ grid, but it does not fill the
whole area: it follows the linear structure of the edges of the
rectilinear embedding.  Therefore, $S$ has $O(n)$ points.  Hence the
triangulation has $O(n)$ edges of length $O(n)$. By calculating all
edge lengths with an absolute error of $O(1/n^2)$, the reduction
algorithm can thus calculate, in polynomial time, a threshold $\hat w$
such that the input formula is satisfiable iff there is a
triangulation of length at most $\hat w$.
\end{proof}

\section{Conclusion}
\label{sec:conclusion}

\subsection{Running Times for Computer Verification of the Proof}
\label{sec:programs}
In designing our gadgets and our proof, we have tried to achieve a
balance between the number of different pieces, which affects the
complexity of the human-readable part of the proof and the number of
case distinctions, and the size of the pieces, which affects the
complexity (running time) for the mechanical part of the proof that
has to be checked by computer (or accepted by faith).

The running time is dominated by the $O(n^3)$ dynamic programming
algorithm for triangulating simple polygons.  The largest point sets
that we handle have 493 points (the left bends).
To total time to run all verifications (with exact integer
interval arithmetic) was about 10 hours on a four-year-old moderate
PC.

\subsection{Dimensions.}


Our gadgets have constant size, but they consist of several
thousands of points and are quite enormous. For example, the clause
gadget (see Figure~\ref{fig:clause}) has dimensions on the order of
$250{\,}000\times 250{\,}000$, with coordinates that are specified as
multiples of $10^{-4}$.  On the other hand, the difference between a
satisfiable and an unsatisfiable SAT instance is reflected in a minute
difference of $0.0007$ in the MWT cost.

With some work, it would be possible to reduce this to some more ``reasonable''
figures or to amplify the difference between satisfiable and
unsatisfiable instances, but we did not find it worth the effort to do
so. First of all, the current gadgets are already a result of tedious
experiments, pushing points into various directions and trying to
understand what happens. Some parts of the design, in particular the
$C$-connections, are very delicate.  Secondly, the dimensions would
still be very large.  One can certainly reduce the constant $230\,000$
of Lemma~\ref{long-distance} by providing a greater variety of
extended wire-pieces, but the bit loop, for example,
(Figure~\ref{variable-loop}) already has size approximately
$7000\times 7000$, and it does not seem easy to push the size very
much below these limits, unless one comes up with a completely
different design.

\subsection{Open Problems}
\label{sec:open}

Several interesting problems remain open.
First of all, it is not known whether the MWT problem is in NP,
since it is not known how to compare sums of Euclidean lengths in polynomial
time %
 \cite{Blomer91},
but this difficulty is more of an algebraic nature.
To define a variant of MWT which is in NP, one can
take the weight of an edge $e$ as the rounded value
$\lceil \Vert e \Vert_2 \rceil$.  
With appropriate scaling, our proof also establishes NP-completeness for
this variant.

Our reduction shows that it is NP-hard to approximate the MWT with a
relative approximation error which is better than $O(1/n^2)$: The
difference between a satisfiable instance of PLANAR 1-IN-3-SAT and an
unsatisfiable instance is reflected in a constant increase of the MWT
cost, and, as mentioned in the proof of~Theorem~\ref{reduction}, the
total cost of the MWT is $O(n^2)$.

One can probably reduce this bound to $O(n\log n)$, and thus establish
that it is NP-hard to achieve a relative approximation error better than
$O(1/(n\log n))$, by using the fact that the interior of a {convex}
$k$-gon of perimeter $p$ can be triangulated with weight $O(p\log k)$.
First, the wires and all gadgets form linear structures of total
length $O(n)$; thus, the length of the gadget boundaries, and the MWT
inside the gadgets is only $O(n)$.  This leaves the holes to be
triangulated.  It should be quite straightforward to extend our
construction in such a way that, apart from $O(n)$ constant-size
holes, only $O(n)$ convex holes with a total of $O(n)$ vertices
remain: one would insert paths of additional points whose
$\beta$-skeleton separates the big holes from the jagged wire
boundaries and cuts the holes into convex pieces, apart from
``linear'' structures that cover only $O(n)$ area.

These non-approximability results do not rule out the existence of a 
polynomial-time approximation scheme.
 For a long time,
attempts to extend techniques from geometric approximation algorithms to the
MWT problem have only led to constant factor approximations
(see \cite{BernEp96} for a survey).
\citeN{RemySt06} showed that it is possible to compute a 
$(1+\varepsilon)$-approximation of the MWT in time $n^{O(\log^8 n)}$,
providing strong evidence that a PTAS might exist. 

In practice, the LMT-skeleton heuristic is extremely fast in computing
LMTs. Combined with with bucketing techniques and fast preprocessing
techniques~\cite{DrysdaleRoAi95}, one empirically achieves almost
linear running times. Thus, in this respect the MWT problem seems to
be similar to the Knapsack Problem, which is also NP-hard but easy to
solve in practice \cite{kpp-kp-07}.  It would be interesting to
analyze the LMT-skeleton heuristic for random point sets.  The good
practical performance indicates that the expected running time
for random inputs might be polynomial, or even close to linear. On the
other hand, as point sets get huge, they will contain, with
non-negligible probability, some larger and larger point
configurations that are hard to triangulate.

\begin{acks}
The authors would like to thank Eric Demaine for pointing out that
the negations in our original paper are not necessary.

The gadgets in this paper were designed using Otfried Cheong's
extensible drawing editor \textsc{ipe} \cite{Schwarzkopf95} as a graphical
user interface. We also used the xml-format of \textsc{ipe} to
generate many of the illustrations of this paper directly
from computer output.

We also thank an anonymous reviewer for extensive remarks that helped to
improve the presentation of the paper.

\end{acks}

\bibliographystyle{acmtrans}
\bibliography{mwt}

\iffull
\appendix

\section{The Data for all Pieces}
\label{appendix-pieces}

\begin{table}[htb]
  \centering
{  

  \offinterlineskip
  \halign{\vrule \strut\hfil\ #\hfil \vrule &\ \hfil $#$ \hfil\vrule
     &&\ \hfil $#$ \hfil\vrule \cr
   \noalign{\hrule}
   \raise2pt\hbox{\strut}pattern & $multiplicity$ &$internal cost $c 
    & $reduced internal cost $\bar c
    & $relative reduced cost $\tilde c\cr\noalign{\hrule}
    \multispan5\vrule\strut\raise3pt\hbox{\strut}\hfil wire-piece\hfil\vrule\cr
    \noalign{\hrule}LL &1& 144.135078832 & 144.135078832 & 0.000000000 \cr
LR &2& 155.655929495 & 144.345585495 & 0.210506663 \cr
RL &1& 132.950328078 & 144.260672078 & 0.125593246 \cr
RR &1& 144.135078832 & 144.135078832 & 0.000000000 \cr
\noalign{\hrule}\noalign{\hrule}
    \multispan5\vrule\strut\raise3pt\hbox{\strut}\hfil extended wire-piece\hfil\vrule\cr
    \noalign{\hrule}LL &1& 455.471523435 & 455.471523435 & 0.000000000 \cr
LR &2& 466.990265006 & 455.679921006 & 0.208397570 \cr
RL &1& 444.283180745 & 455.593524745 & 0.122001310 \cr
RR &1& 455.471523435 & 455.471523435 & 0.000000000 \cr
\noalign{\hrule}\noalign{\hrule}
    \multispan5\vrule\strut\raise3pt\hbox{\strut}\hfil thickening adapter\hfil\vrule\cr
    \noalign{\hrule}LL &1& 7188.980062460 & 7207.382390460 & 0.000051402 \cr
LR &1& 7237.113898304 & 7207.401226304 & 0.018887246 \cr
RL &1& 7177.684294307 & 7207.396966307 & 0.014627250 \cr
RR &1& 7225.784667058 & 7207.382339058 & 0.000000000 \cr
\noalign{\hrule}\noalign{\hrule}
    \multispan5\vrule\strut\raise3pt\hbox{\strut}\hfil $C$ connection\hfil\vrule\cr
    \noalign{\hrule}LLl &1& 14027.752986494 & 14033.408158494 & 0.003861076 \cr
LLr &1& 14039.059470993 & 14033.404298993 & 0.000001575 \cr
LRl &1& 14075.894120134 & 14033.434292134 & 0.029994716 \cr
LRr &1& 14087.200604634 & 14033.430432634 & 0.026135215 \cr
RLl &1& 13979.654697176 & 14033.424869176 & 0.020571757 \cr
RLr &1& 13990.965042750 & 14033.424870750 & 0.020573332 \cr
RRl &1& 14027.749125419 & 14033.404297419 & 0.000000000 \cr
RRr &1& 14039.064100296 & 14033.408928296 & 0.004630878 \cr
\noalign{\hrule}\noalign{\hrule}
    \multispan5\vrule\strut\raise3pt\hbox{\strut}\hfil $C_0$ connection\hfil\vrule\cr
    \noalign{\hrule}LL &1& 12733.864882577 & 12733.864882577 & 0.000000000 \cr
LR &2& 12782.023884279 & 12733.908884279 & 0.044001701 \cr
RL &2& 12685.770454334 & 12733.885454334 & 0.020571757 \cr
RR &1& 12733.864882577 & 12733.864882577 & 0.000000000 \cr
\noalign{\hrule}\noalign{\hrule}
    \multispan5\vrule\strut\raise3pt\hbox{\strut}\hfil left bend\hfil\vrule\cr
    \noalign{\hrule}LL &1& 5425.386907832 & 5425.386907832 & 0.000000000 \cr
LR &2& 5436.783712726 & 5425.473368726 & 0.086460895 \cr
RL &2& 5414.202157077 & 5425.512501077 & 0.125593246 \cr
RR &1& 5425.386907832 & 5425.386907832 & 0.000000000 \cr
\noalign{\hrule}\noalign{\hrule}
    \multispan5\vrule\strut\raise3pt\hbox{\strut}\hfil right bend\hfil\vrule\cr
    \noalign{\hrule}LL &1& 5271.525633165 & 5271.525633165 & 0.000000000 \cr
LR &2& 5283.046483828 & 5271.736139828 & 0.210506663 \cr
RL &2& 5260.340882410 & 5271.651226410 & 0.125593246 \cr
RR &1& 5271.525633165 & 5271.525633165 & 0.000000000 \cr
\noalign{\hrule}\noalign{\hrule}
    \multispan5\vrule\strut\raise3pt\hbox{\strut}\hfil thick left bend\hfil\vrule\cr
    \noalign{\hrule}LL &1& 23329.163703675 & 23329.163703675 & 0.000000000 \cr
LR &2& 23378.169964722 & 23330.054964722 & 0.891261046 \cr
RL &2& 23281.069275432 & 23329.184275432 & 0.020571757 \cr
RR &1& 23329.163703675 & 23329.163703675 & 0.000000000 \cr
\noalign{\hrule}
}

}
  \caption{Analysis of the pieces}
  \label{tab:pieces}
\end{table}

Table~\ref{tab:pieces} shows, for all pieces, the internal cost,
the reduced cost, 
 and the relative reduced cost, 
for all configurations of boundary triangles.
Figures~\ref{fig:TABLE-C0}--\ref{fig:TABLE-thick-left-bend},
as well as Figure~\ref{fig:TABLE-wire}, 
show for each configuration an optimal triangulation.
The two resizing wire-pieces are symmetric, and thus only one of them,
the thickening adapter, is given. When applying the results to the
thinning adapter, the labels must be flipped: LL becomes RR and RR
becomes LL.
The raw costs for the thick left bend are the same as for the 
left bend, scaled by~4.3. But since the large terminal triangles use a
different value of $\delta$, the reduced costs bear no direct relation
to those of the left bend.
The last column is what is summarized in Table~\ref{tab:pieces-summary}.

We extended our \textsc{Python} programs to output their results in the form of
tables (in \TeX\ format) and figures.  As the output format for the
figures, we used the xml-format of the \textsc{Ipe} program\footnote
{\url{http://tclab.kaist.ac.kr/ipe/}} \cite{Schwarzkopf95}, which can be
directly converted to eps or pdf format or further edited.  Many
illustrations of this paper were obtained in this way.  In particular,
the table and all figures of this section, as well as
Figures~\ref{fig:TABLE-wire} 
and Table~\ref{W-lemma}, were generated directly from our 
programs without manual intervention.

\begin{figure}[htbp]
  \centering
\includegraphics[width=\textwidth]{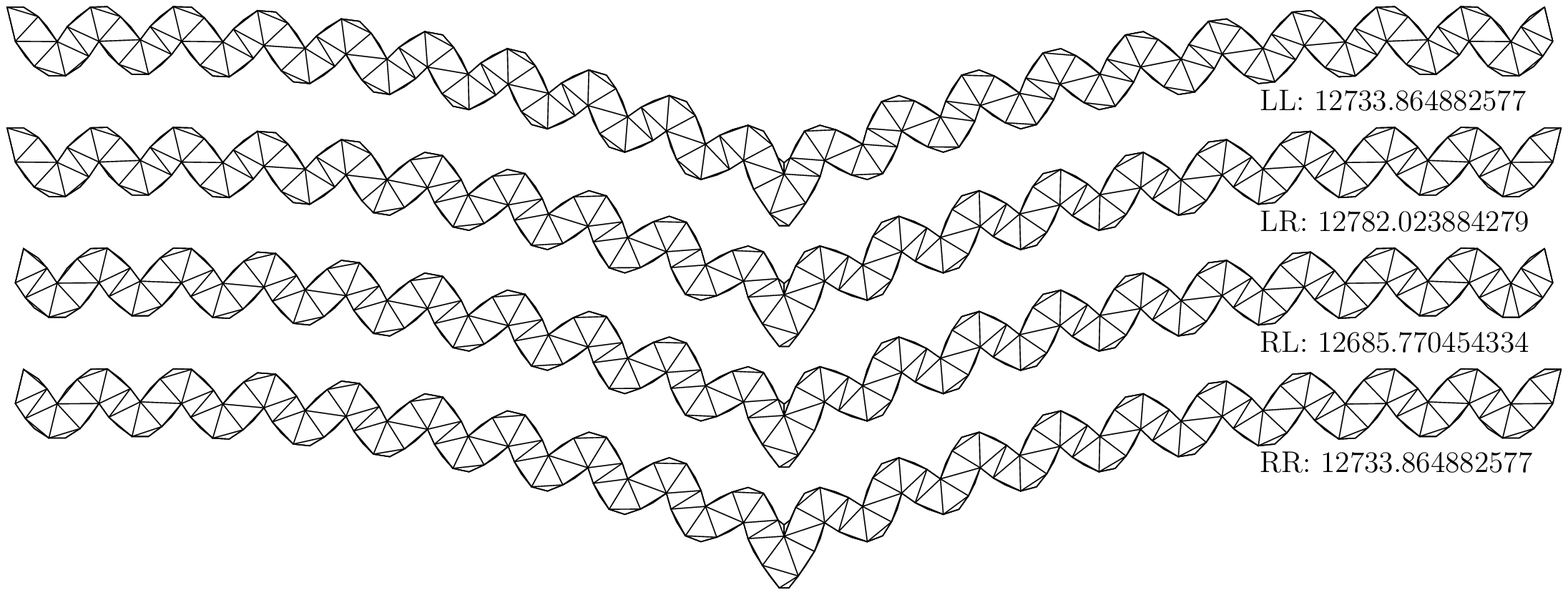}
  \caption{Optimal solutions for all cases for the $C_0$ connection}
  \label{fig:TABLE-C0}
\end{figure}

\begin{figure}[htbp]
  \centering
\includegraphics[width=\textwidth]{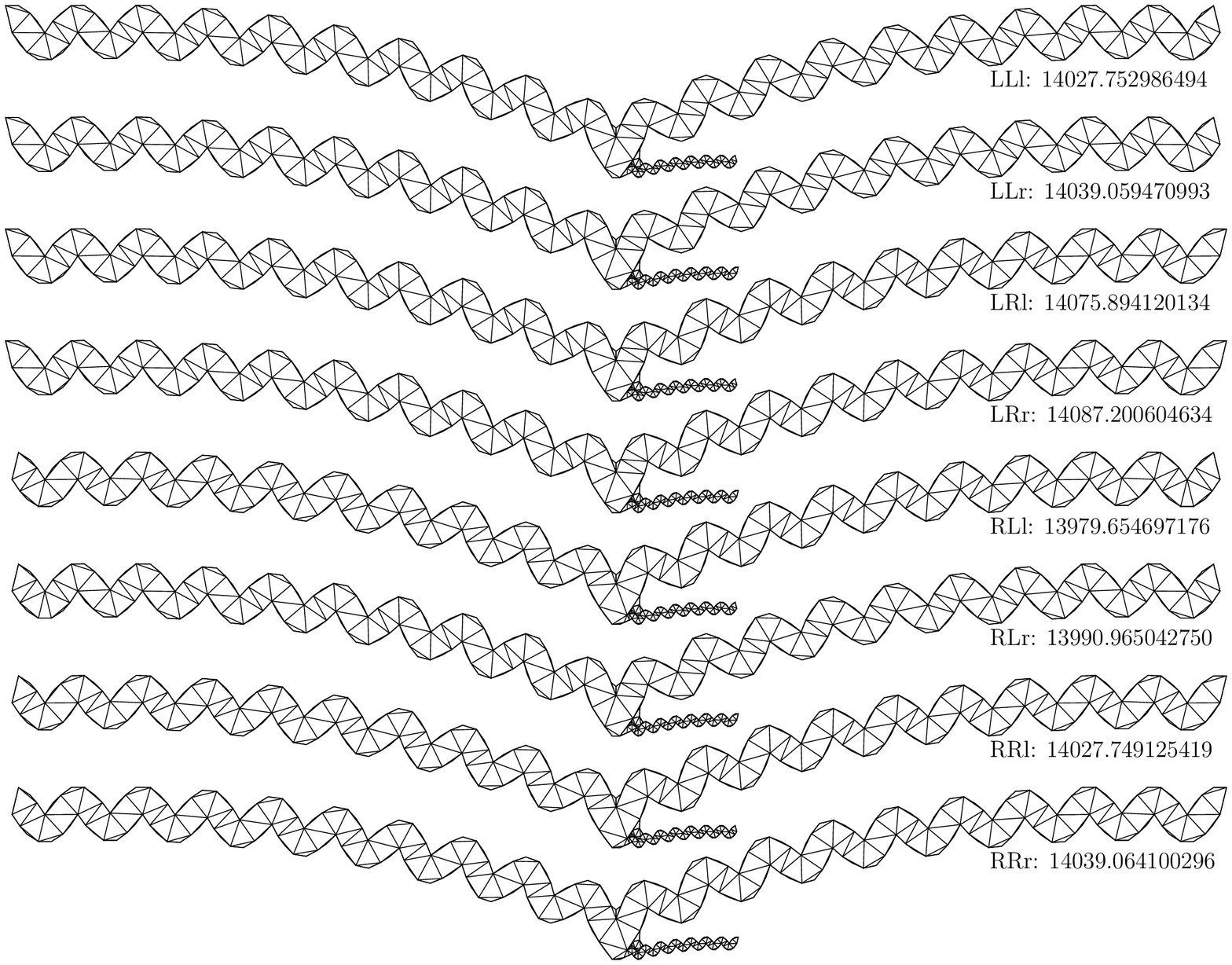}
  \caption{Optimal solutions for all cases for the $C$ connection}
  \label{fig:TABLE-C}
\end{figure}

\begin{figure}[htbp]
  \centering
\includegraphics[width=\textwidth]{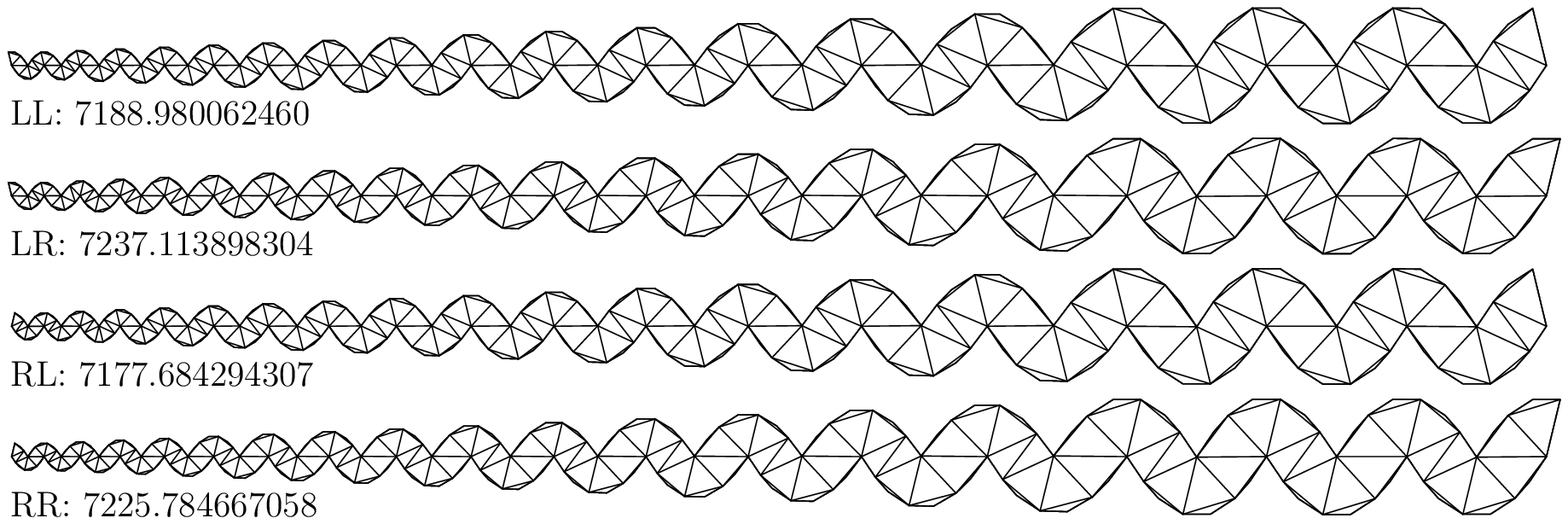}
  \caption{Optimal solutions for all cases for the thickening adapter}
  \label{fig:TABLE-resize}
\end{figure}

\begin{figure}[htbp]
  \centering
\includegraphics[width=\textwidth]{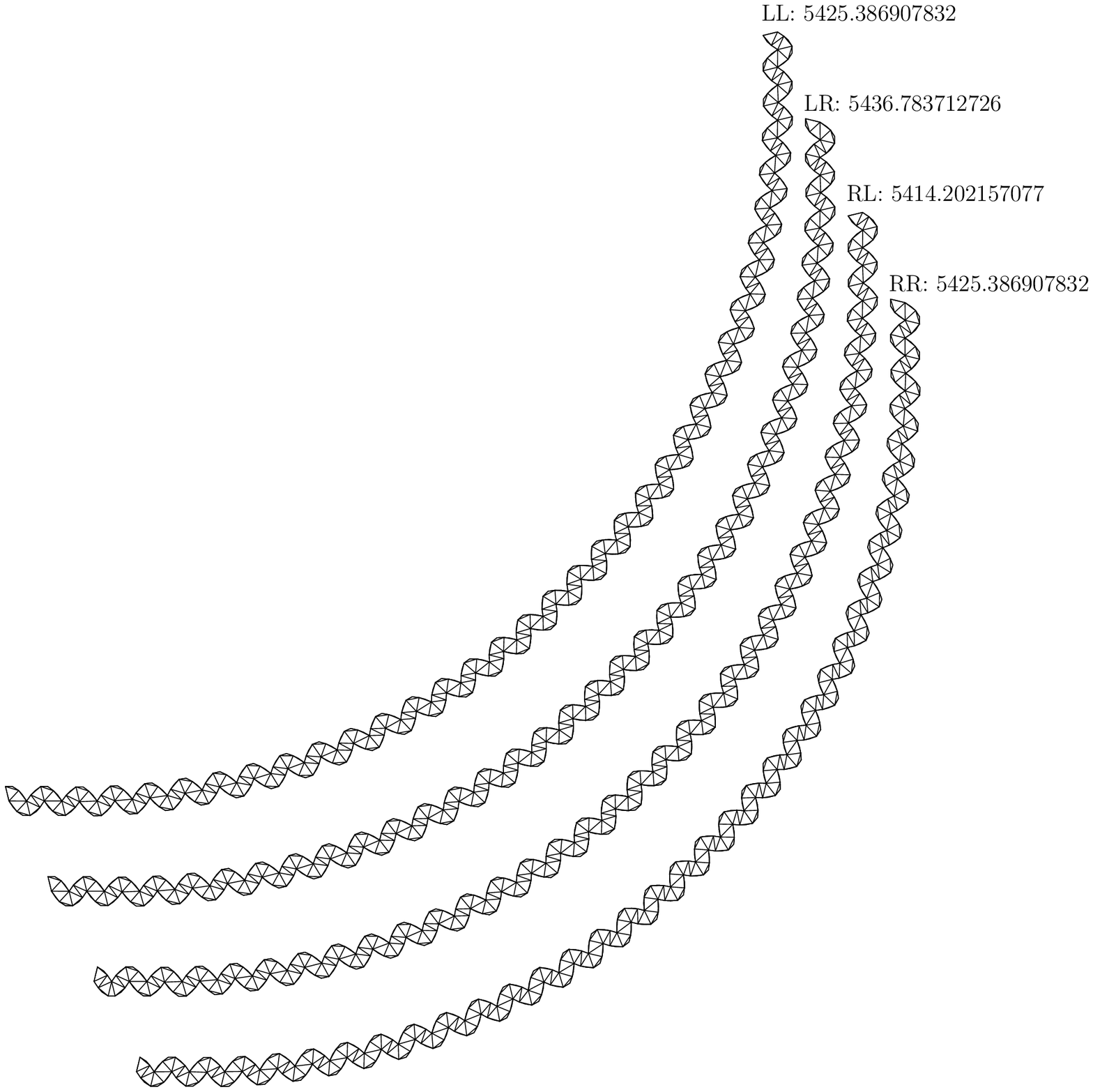}
  \caption{Optimal solutions for all cases for the left bend}
  \label{fig:TABLE-left-bend}
\end{figure}

\begin{figure}[htbp]
  \centering
\includegraphics[width=\textwidth]{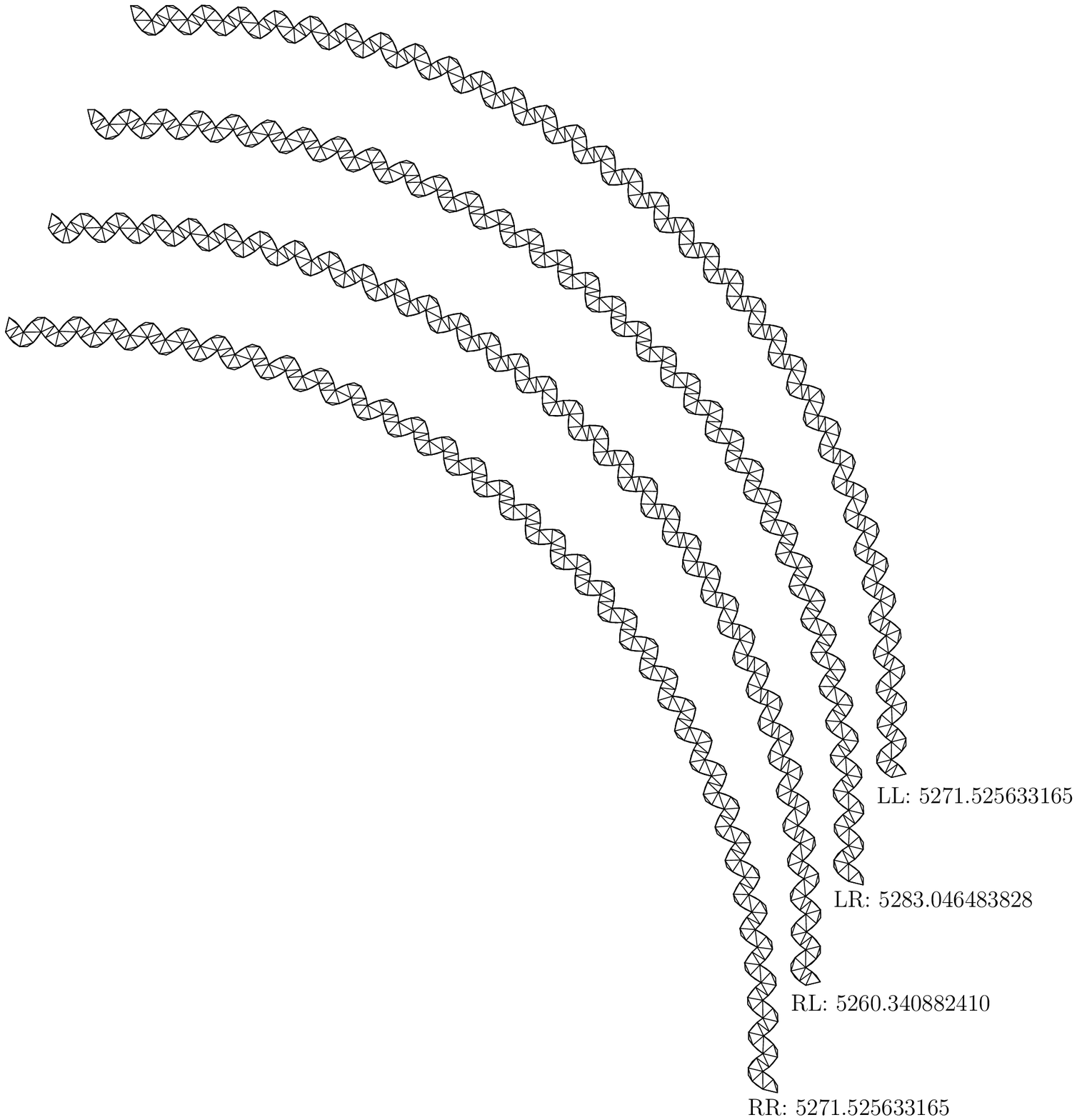}
  \caption{Optimal solutions for all cases for the right bend}
  \label{fig:TABLE-right-bend}
\end{figure}

\begin{figure}[htbp]
  \centering
\includegraphics[width=\textwidth]{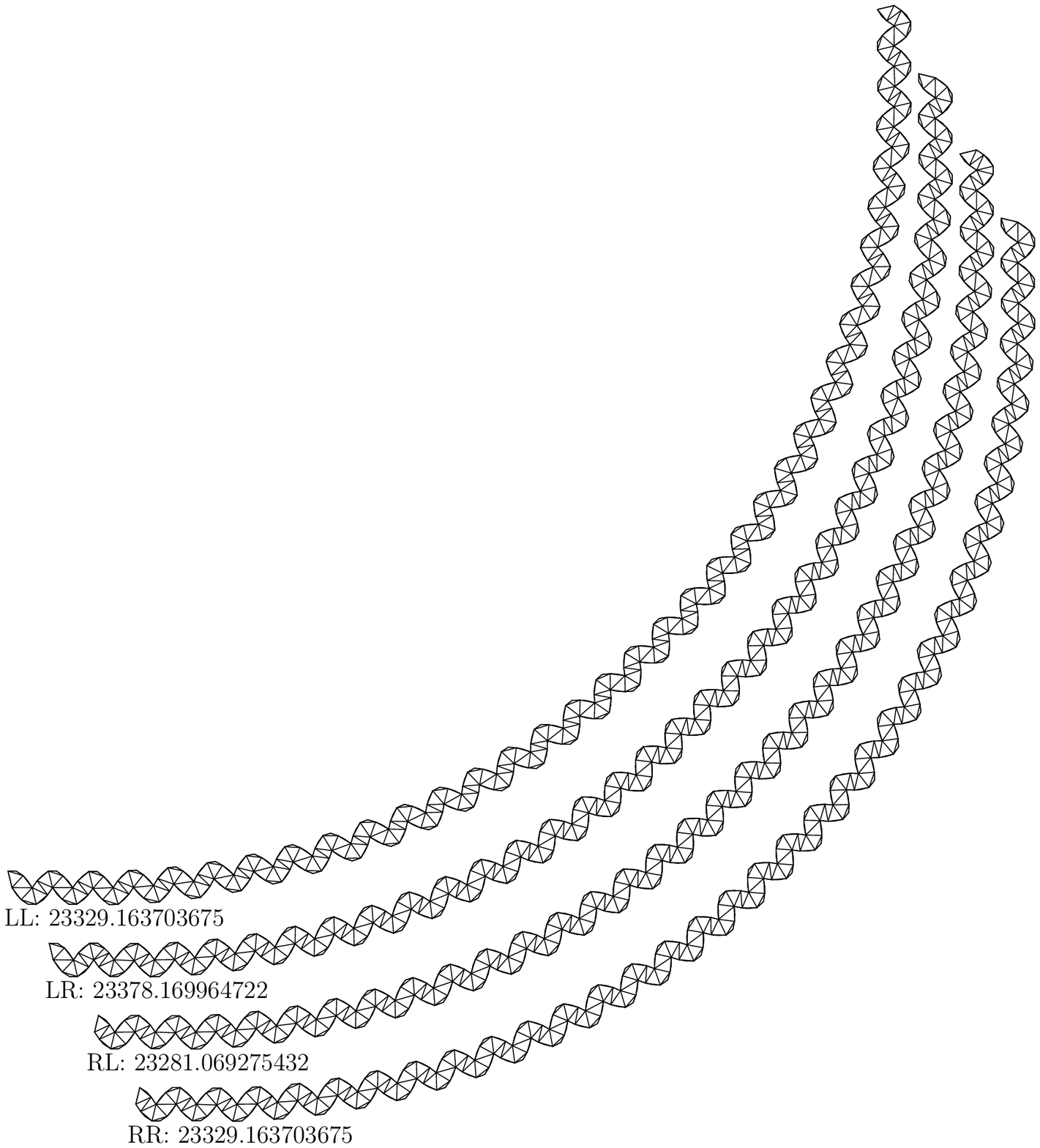}
  \caption{Optimal solutions for all cases for the thick left bend}
  \label{fig:TABLE-thick-left-bend}
\end{figure}

\section{Computer Programs}
\label{app:programs}

We have written programs that check all of the claimed properties.
(Propositions~\ref{basic-properties}--\ref{boundary}).
If any condition fails, the programs raise an exception.
To give an idea of what was automatically checked, we
show an excerpt of the log-file, concerning the
extended wire-piece (Figures~\ref{definition-wire-extended}
and~\ref{fig:TABLE-wire}).
{\small
\begin{verbatim}
======================== extended wire-piece ====================
All coordinates are multiples of 0.0001
terminal triangle basepoint: (-41.100,0)
     edge vector in state L: (-2.700,11.2)
     edge vector in state R: (2.700,11.2)
terminal triangle basepoint: (41.110,0)
     edge vector in state L: (-2.700,11.2)
     edge vector in state R: (2.700,11.2)
All terminal coordinates are multiples of 0.01
0 duplicate point(s).
The point set is symmetric with respect to the vertical axis x=0.005.
cos(alpha)^2 = 3120343/10000000 = 0.312034; beta = 1.205637.
case LL: 44 points. 455.471523435 455.471523435
case LR: 45 points. 466.990265006 455.679921006
case RL: 43 points. 444.283180745 455.593524745
case RR: 44 points. 455.471523435 455.471523435
\end{verbatim}
}
The complete \textsc{Python} source code, as well as the data for the pieces, are available
on the Internet.%
\footnote{%
  \url{http://www.inf.fu-berlin.de/inst/ag-ti/people/rote/Software/MWT/python/}}
In total, there are about 1000 lines of code.
The programs are mostly rather straightforward, so we only show
four modules:
The module \verb|mwt_polygon.py| contains the dynamic programming
algorithm for computing the MWT of a simple polygon
 (Section~\ref{app:dynprog}).
It relies on
the module \verb|arithmetic.py| for the representation of
fixed-precision decimal quantities as (long) integers, for
(rudimentary) interval arithmetic with integers, and for calculating
the Euclidean length as an interval (Section~\ref{app:arithmetic}).
The module \verb|check_beta.py| determines the smallest value $\beta$ for which the given boundary edges belong to the $\beta$-skeleton (Section~\ref{app:beta-skeleton}).
The module \verb|check_W.py| 
checks all cases for the point set $W$ for proving
Lemma~\ref{lemma-two-edges} (Section~\ref{app:two-edges}).

\subsection{Arithmetic}
\label{app:arithmetic}

Since the dynamic programming algorithm involves only additions and
comparisons, interval arithmetic is not a big deal.  We use
fixed-point arithmetic, which is simulated by integer
arithmetic. Hence all calculations are exact, and we need not care
about directed rounding when doing arithmetic with intervals.

The only non-trivial part is the calculation of Euclidean distances
by the \texttt{length} function,
which involves a square root.  We use the \texttt{decimal} package of
\textsc{Python}, which provides an arbitrary-precision square root
operation, to obtain a starting approximation of the square root.  The
approximation is then checked and refined directly, in a
straightforward way.

{\small
\begin{verbatim}
from decimal import *

cdigits = 5
scale = 100000 # initial values

def setprecision(computedigits):
  global cdigits,scale
  cdigits = computedigits
  scale = 10**computedigits

def length (p1,p2):
    """Euclidean distance between two points
    p1 and p2 are points with integer coordinates.
    The results is an integer interval."""
    l_squared = (p2[0]-p1[0])**2 + (p2[1]-p1[1])**2

    # obtain an initial approximation using the
    # square root operation of the decimal package
    oldprec = getcontext().prec
    getcontext().prec = len(str(l_squared))/2+2
    a = int(Decimal(l_squared).sqrt().quantize(Decimal(1),
                                               rounding=ROUND_FLOOR))
    getcontext().prec = oldprec # restore the old precision

    # correct and verify the result.
    # The final result will be correct regardless of how good
    # the initial approximation was (although it may take longer).
    while a**2 > l_squared:      a -= 1
    while (a+1)**2 <= l_squared: a += 1
    if a*2 < l_squared: return Interval_integer(a,a+1)
    else:               return Interval_integer(a,a)
  
def decimal_to_scaled_int(s):
    return int(s*scale)
  
def print_scaled_int(n,ndigits=None):
    """print integer with decimal point inserted
       "ndigits" digits from the right end"""
    if n<0:
        sign = "-"
        n = -n
    else:
        sign = ''
    if ndigits==None: ndigits=cdigits # default
    return sign + ("%d.%0"+str(ndigits)+"d") % divmod(n, 10**ndigits)

class Interval_integer(object):
  """ represents an interval [a,b] with a<=b """
  def __init__(self,a,b=None):
    self.a = a
    if b==None:
        self.b = a
    else:
        self.b = b

  def __add__(self,y):
    if isinstance(y, (int,long)):
      return Interval_integer(self.a + y, self.b + y)
    else: # assume y is an Interval_integer:
      return Interval_integer(self.a + y.a, self.b + y.b)

  def quantize(self,ndigits):
    """try to find a unique value which is guaranteed
    to be the rounded value of the interval"""
    scalehalf = 10**ndigits/2
    rounded_value = (self.a/scalehalf + 1)/2
    val = rounded_value*2*scalehalf
    if val-scalehalf < self.a and val+scalehalf > self.b:
        return rounded_value
    print self, "cannot be rounded uniquely to",
    print ndigits, "fewer digits."
    raise ValueError

  def quantize_and_print(self,ndigits,initialdigits=None):
    """print the unique value which is the rounded value
    of the interval with "ndigits" remaining digits, if it exists"""
    if initialdigits==None: initialdigits=cdigits # default
    return print_scaled_int(
                  self.quantize(initialdigits-ndigits), ndigits)
  
  def __str__(self):
    return "["+str(self.a)+","+str(self.b)+"]"
  def __repr__(self):
    return "Interval("+repr(self.a)+","+repr(self.b)+")"

def keep_best(l):
  """keeps a list that is guaranteed to contain the smallest
     value from a list of intervals"""
  threshold = min(x.b for x in l)
  return [ x for x in l if x.a <= threshold ]

\end{verbatim}
}
\subsection{Dynamic Programming for Triangulating a Polygon}
\label{app:dynprog}

The procedure \texttt{mwt} of the module \verb|mwt_polygon.py|
implements the classical $O(n^3)$ dynamic programming algorithm for
optimally triangulating a simple polygon $p$ \cite{Gilbert79,Klincsek80}. It
has an optional argument \texttt{excluded} for specifying edges that
cannot be used in the triangulation.  The input polygon $p$ is assumed
to have integer coordinates.  Euclidean edge lengths are calculated as
integer intervals using \texttt{arithmetic.length()}.

In contrast to \citeN{Klincsek80}, our
 procedure does not explicitly test the edges for crossings; it
only tests whether all triangles that are used in the triangulation are
oriented counterclockwise. If the input is a simple polygon, this is
sufficient to ensure that the resulting triangulation is non-crossing,
see Lemma~\ref{triangulate-simple-polygon} below.
(Hence, the program will also triangulate certain non-simple
polygons, but it will not triangulate polygons which are oriented
clockwise!)

To show that the our constructions do not depend on assumptions about
handling point sets which are not in general position, triangles with
three points on a line are (temporarily) considered as valid triangles
of a triangulation, but these triangles, as well as any (partial)
triangulations that contain such triangles, are flagged as degenerate.
If such a triangulation would ``survive'' as a candidate for an optimal solution, the checking routines would report an error
(see for example the procedure \verb|check_W_cases()| in
Section~\ref{app:two-edges}).
Triangles with coinciding vertices are not considered.

\begin{lemma}\label{triangulate-simple-polygon}
  Let $T$ be a set of triplets $(i,j,k)$ with $1\le i<j<k\le n$ such
  that for a convex polygon $P=p_1\ldots p_n$, the triangles
  $p_ip_jp_k$ for $(i,j,k)\in T$
  form a triangulation of $P$.

  Let $P$ be an arbitrary simple polygon.  If all triangles
  $p_ip_jp_k$ for $(i,j,k)\in T$ are oriented counter-clockwise, they
  form a triangulation of~$P$.
\end{lemma}
\begin{proof}
  This can be seen by counting the number of triangles in which a
  given point $x$ of the plane is contained. This number can only
  change when $x$ crosses an edge of a triangle. However, all triangle
  edges have another triangle on the opposite side, with the exception
  of the triangle edges that are edges of $P$.  Thus, the number of
  triangles in which a point of the plane is contained is constant
  except at the boundary of~$P$, where it changes by $\pm1$. Since
  this number is 0 when the point $x$ is far away, every point $x$ in
  the interior of $P$ is covered by exactly one triangle, and no
  triangle sticks out of~$P$.
\end{proof}
{\small
\begin{verbatim}
import arithmetic

def mwt(p,excluded=[]):
  """compute minimum-weight triangulation of a simple polygon p
     by dynamic programming, using integer interval arithmetic.
     The coordinates of p are integers.

     Some edges may be excluded from consideration.
     The polygon p must be oriented counter-clockwise!
     Otherwise, no solution will be found.
     The result is a list of triplets
          (weight, solution, degenerate_flag)
     where weight is given as an interval, sorted by the
     lower bound weight.a of the solution quality
     To show that the result does not depend on assumptions
     about handling point sets which are not in general position,
     degenerate solutions (triangles with three points on a line)
     are considered, but they are flagged as degenerate."""
  opt = {} ## array of optimal (partial) solutions.
  n = len(p)
  for i in range(n-1): # initialization
    o = arithmetic.Interval_integer(0)
             ## calculate the weight only of interior edges
    o.info = None # backtracking information and flag is stored
                  # within the Interval_integer objects
    o.degenerate_flag = False
    opt[i,i+1] = [o]
    # opt[i,j] stores a LIST of all potential candidate solutions.
    
  for l in range(2,n): # l="length" of the edges
    for i in range(n-l):
      j = i+l
      # check for triangulations containing edge (i,j):
      if (i,j) in excluded or (j,i) in excluded:
        ## we are interested in MWT that does not contain those edges
        continue
      if p[i]==p[j]: # this point appears twice on the boundary
        continue
      if (i,j) == (0,n-1):
        edgelength = 0 # final edge is not counted
      else:
        edgelength = arithmetic.length(p[i],p[j])
      for k in range(i+1,j):
        if opt.has_key((i,k)) and opt.has_key((k,j)):
          ori = orientation(p[i],p[k],p[j])
          if ori >= 0:
            in1 = opt[i,k]
            in2 = opt[k,j]
            allcand = [ computesol(in1,in2,n1,n2,k,ori==0,edgelength)
                          for n1 in range(len(in1))
                          for n2 in range(len(in2)) ]
            opt[i,j] = arithmetic.keep_best(opt.get((i,j),[])+allcand)
  # sort final result by lower bounds of intervals:
  allopt = [(x.a,x) for x in opt[0,n-1]]
  allopt.sort()
  opt[0,n-1] = [x[1] for x in allopt]
  # retrace optimal solution:
  return [ (arithmetic.Interval_integer(x.a,x.b),
                       retrace(opt,0,n-1,ind), x.degenerate_flag)
           for x,ind in zip( opt[0,n-1], range(len(opt[0,n-1])) ) ]

def orientation(p1,p2,p3):
  x2 = p2[0]-p1[0]
  y2 = p2[1]-p1[1]
  x3 = p3[0]-p1[0]
  y3 = p3[1]-p1[1]
  return x2*y3-x3*y2

def computesol(l1,l2,i1,i2,k,flag,edgelength):
    r = l1[i1] + l2[i2] + edgelength
    r.info = (k,i1,i2) # k=splitting point.
                       # i1,i2: pointers into sublists
    r.degenerate_flag = (flag or l1[i1].degenerate_flag
                              or l2[i2].degenerate_flag)
    return r

def retrace(opt,i,j,ind):
  """recursively recover the solution, based on the "info"
     fields stored with the solution values in the array "opt".
     Recover the solution for position "ind" in the list "opt".
     The result is a list of triplets (i,k,j), each representing
     the two edges p[i]p[k] and p[k]p[j]."""
  if opt[i,j][ind].info:
    k,i1,i2 = opt[i,j][ind].info
    return [(i,k,j)]+retrace(opt,i,k,i1)+retrace(opt,k,j,i2)
  else:
    return []
\end{verbatim}
}



\subsection{Checking the $\beta$-skeleton}
\label{app:beta-skeleton}
This procedure is straightforward: for every edge $pq$ of the
boundary, it runs through all remaining points $r$ and checks whether
they violate the $\beta$-skeleton condition, by calculating the
(squared) cosine of the angle $\alpha=prq$ with the cosine law.  The
running time is $O(n)$ per edge of the boundary, thus at most $O(n^2)$
in total. The relation between the angle $\alpha$ and the
ratio $\beta$ between the diameter of the circumcircle of $pqr$ and the
distance $|pq|$ is given by
$$
\sin\alpha = 1/\beta
\text{, \ or \ }
\cos^2\alpha = 1-1/\beta^2
$$

To understand this program, one has to know how the pieces are
given. Each \texttt{piece} is represented as a dictionary.
The field
\texttt{piece["partlist"]} is a sequence of (names of) boundary parts,
usually just \texttt{("lower", "upper")};
 only the $C$-connection has three parts.
 Then \texttt{piece["lower"]} and \texttt{piece["upper"]} (or whatever
 the names in the \texttt{"partlist"} are) are the actual boundary
 parts, as lists of points. Each point is a pair of precise
 \texttt{Decimal} numbers as provided by the \texttt{decimal} package.

{\small
\begin{verbatim}
from math import sqrt
from checking_routines import smallest_unit
from basic_routines import map_coordinates

beta_threshold = 1.18  # ">=1.17683 !"
betaprecision = 10**7

def coordinate_to_int(c):
    "scale coordinate to integer"
    return int(c/smallest_unit) # smallest_unit=0.0001

def length2(p):
    """squared length of vector"""
    return p[0]**2 + p[1]**2

def checkbeta(piece):
  min_cos_squared = 100*betaprecision # infinity
  for part in piece["partlist"]:
    # scaling all coordinates does not affect beta-skeleton:
    poly = map_coordinates(coordinate_to_int, piece[part])
    for i in range(len(poly)-1): # run through all edges:
      p1 = poly[i]
      p2 = poly[i+1]
      # check edge (p1,p2):
      # run through all points p: 
      for part_2 in piece["partlist"]:
        for p in map_coordinates(coordinate_to_int,piece[part_2]):
          if p<>p1 and p<>p2:
            # compute cos_squared, representing the
            # squared cosine of the angle alpha=p1,p,p2
            x1 = (p1[0]-p[0], p1[1]-p[1])
            x2 = (p2[0]-p[0], p2[1]-p[1])
            innerproduct = x1[0]*x2[0] + x1[1]*x2[1]
            # so far, all calculations were exact.
            if innerproduct>0:
               cos_squared = ( innerproduct**2 * betaprecision /
                                      (length2(x1)*length2(x2)) )
               # scaled by betaprecision and rounded down;
               # yields a conservative (lower) estimate of beta
            min_cos_squared = min(min_cos_squared,cos_squared)
  min_cos_squared_unscaled = min_cos_squared/float(betaprecision)
  beta = 1/sqrt(1-min_cos_squared_unscaled)
  # This final calculation is just in double-precision floating-point.
  print "cos(alpha)^2 = %d/%d = %8.6f; beta = %8.6f." % (
    min_cos_squared, betaprecision, min_cos_squared_unscaled, beta)
  if beta < beta_threshold:
    print "beta too small: ", beta, ">=1.17683 !", beta_threshold
    raise RuntimeError
  return beta
\end{verbatim}
}
\subsection{Proving Lemma~\ref{lemma-two-edges}}
\label{app:two-edges}
This program runs through all 21 subpolygons of the polygon $W$ in the proof
of Lemma~\ref{lemma-two-edges}. For illustration, we give here
the output produced for the first case:
{\small
\begin{verbatim}
Case v1, v1': difference =
4.00304
          Best solution value without terminal edges:
              [202.72577,202.72594]
          There is/are 2 best solution(s) without restriction:
              [198.72255,198.72273]
              [198.72255,198.72273]
\end{verbatim}
}
There is an extended version (not shown here) which also generates a
drawing of the triangulations, and which has been used to generate
Table~\ref{W-lemma}.  The module \verb|solve_all_patterns.py|,
 by which the data of
Appendix~\ref{appendix-pieces} were generated, is similar.

{\small
\begin{verbatim}
# checks presence of one ot two "terminal" edges in thin wire

from basic_routines import map_coordinates
import arithmetic
from arithmetic import print_scaled_int, decimal_to_scaled_int
import mwt_polygon

from pieces import W_upper, W_lower, W_point_x, W_point_y, W_point_z
# W_upper and W_lower form the set W.

def check_W_cases():
   mindiff = 999 * arithmetic.scale # infinity

   for i in range(1,7):
     for j in range(i,7): #  for 1<=i<=j<=6:

       p = W_lower[i:len(W_lower)-j] + W_upper
       i_x = p.index(W_point_x)
       i_y = p.index(W_point_y)
       i_z = p.index(W_point_z)
       excl = [(i_x,i_y),(i_x,i_z)] # critical edges

       pcalc = map_coordinates(decimal_to_scaled_int, p)

       # First, compute optimum solution without restriction
       elist = mwt_polygon.mwt(pcalc)

       for (result,sol,degenerate) in elist:
          # convert list of triplets to list
          # of pairs representing edges:
          sole = [(x,y) for [x,y,z] in sol] + \
                 [(y,z) for [x,y,z] in sol]
          if not((i_x,i_y) in sole or (i_x,i_z) in sole or
                 (i_y,i_x) in sole or (i_z,i_x) in sole):
              print "incorrect:",i,j,result,sol,degenerate
              print ("optimal solution does not contain" ,
                        i_x,i_y, "or", i_x,i_z)
              raise StandardError
          if degenerate:
              print "DEGENERATE SOLUTION:",i,j,result,sol,degenerate
              raise StandardError

       # Now, compute optimum solution with the two edges excluded:
       elist2 = mwt_polygon.mwt(pcalc, excluded=excl)

       # consider just the first solution from the list:
       diff = elist2[0][0].a - max(result.b for (result,s,f) in elist)
       mindiff = min(mindiff,diff)

       print "Case v%d, v%d': difference =" % (i,j)
       print print_scaled_int(diff)
       print " "*10+"Best solution value without terminal edges:"
       print_scaled_interval(elist2[0][0])
       print "          There is/are",len(elist),
       print    "best solution(s) without restriction:"
       for result,sol,f in elist: print_scaled_interval(result)
       print
   print "Minimum overall difference =", print_scaled_int(mindiff)

def print_scaled_interval(x):
 print " "*14+"[%s,%s]"%(print_scaled_int(x.a), print_scaled_int(x.b))

print "Program check_W.py. Check terminal edges xy and xz in W"
arithmetic.setprecision(5) # 5 digits are sufficient.
check_W_cases()
\end{verbatim}
}


\vbox{\hbox to 0pt{\hss \vbox to 0pt{\vss
\endreceived
\hrule height 0pt
}}}
\newcommand\touch[1]{{\setbox0=\vbox{\verbatiminput{python/#1}}}}
\def\touch#1 {{\setbox0=\vbox{\hsize=50cm\verbatiminput{python/#1}}}}
\touch arithmetic.py
\touch basic_routines.py
\touch check_W.py
\touch check_W_with_figures.py
\touch check_beta.py
\touch checking_routines.py
\touch mwt_polygon.py
\touch output_routines.py
\touch pieces.py
\touch pieces_data.py
\touch run_all_pieces.py
\touch solve_all_patterns.py
\touch makeeps-files
\touch makepdf-files
\touch results/logfile
\touch results/W-logfile
\touch results/all_pieces_coordinates.txt
\touch README.txt
\fi
\end{document}